%% file: main.tex
\documentclass[lettersize,journal]{IEEEtran}
\ifCLASSOPTIONcompsoc
  \usepackage[nocompress]{cite}
\else
  \usepackage{cite}
\fi
\ifCLASSINFOpdf
   \usepackage[pdftex]{graphicx}
\else
\fi
\usepackage[hyphens]{url}

\usepackage{xcolor}
\usepackage{booktabs}
\usepackage{hyperref}
\usepackage{cleveref}
\usepackage{caption}
\usepackage{subcaption}
\usepackage{lipsum}
\usepackage{multirow}
\usepackage{stfloats}

\newcounter{mainfindingid}
\newcommand{\defmainfinding}[2]{\refstepcounter{mainfindingid}\label{#1}\item[O-\arabic{mainfindingid}:] #2}
\newcommand{\refmainfinding}[1]{\textbf{O-\ref{#1}}}

\begin{document}
%
\title{Cloud Uptime Archive: Open-Access Availability Data of Web, Cloud, and Gaming Services}
\author{Sacheendra Talluri, Dante Niewenhuis, Xiaoyu Chu, Jakob Kyselica, Mehmet Cetin, Alexander Balgavy, Alexandru Iosup}%
\IEEEtitleabstractindextext{%
\begin{abstract}
\input{sections/0.abstract}
\end{abstract}

}

\maketitle

\IEEEdisplaynontitleabstractindextext

%
\IEEEpeerreviewmaketitle

\input{sections/1.intro}

\input{sections/2.sys_model}

\input{sections/3.selection_and_validation}
\input{sections/4.mtbf_and_mttr}

\input{sections/5.time_pattern}

\input{sections/6.severity}

\input{sections/7.causes}
\input{sections/8.checkpointing}

\input{sections/8.experiment}
\input{sections/9.conclusion}


%





\ifCLASSOPTIONcaptionsoff
  \newpage
\fi



\bibliographystyle{IEEEtran}
\bibliography{bibliography}

%

\begin{IEEEbiography}[{\includegraphics[width=1in,height=1.25in,clip,keepaspectratio]{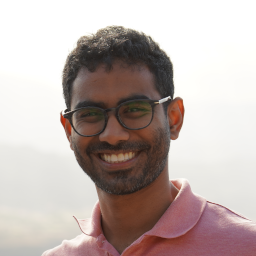}}]{Sacheendra Talluri} is a PhD student at Vrije Universiteit Amsterdam. His work focuses on fault tolerance in the cloud and the structure of cloud resource managers.
\end{IEEEbiography}

\begin{IEEEbiography}[{\includegraphics[width=1in,height=1.25in,clip,keepaspectratio]{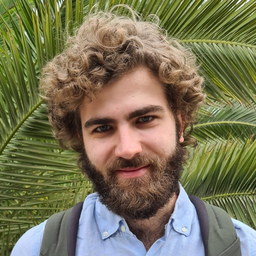}}]{Dante Niewenhuis}
is a PhD student at Vrije Universiteit Amsterdam. His work focuses on using simulation and digital twinning to improve the sustainability of datacenters.
\end{IEEEbiography}

\begin{IEEEbiography}[{\includegraphics[width=1in,height=1.25in,clip,keepaspectratio]{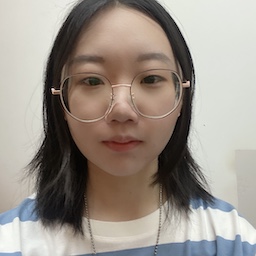}}]{Xiaoyu Chu}
is a PhD student at Vrije Universiteit Amsterdam. Her work characterizes the reliability of HPC datacenters and popular ML-focused cloud services.
\end{IEEEbiography}

\begin{IEEEbiography}[{\includegraphics[width=1in,height=1.25in,clip,keepaspectratio]{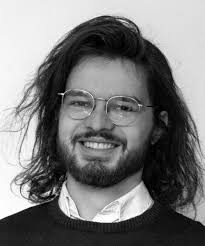}}]{Jakob Kyselica}
is a Software Engineer at MEOS. He obtained his MSc from Vrije Universiteit Amsterdam focusing on the reliability of online games.
\end{IEEEbiography}

\begin{IEEEbiography}[{\includegraphics[width=1in,height=1.25in,clip,keepaspectratio]{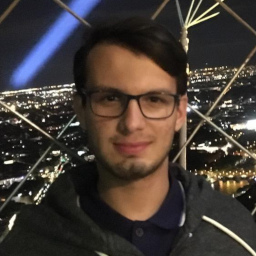}}]{Mehmet Cetin}
is a Software Engineer at AMazon. He obtained his MSc from Vrije Universiteit Amsterdam focusing on the reliability of public cloud services.
\end{IEEEbiography}

\begin{IEEEbiography}[{\includegraphics[width=1in,height=1.25in,clip,keepaspectratio]{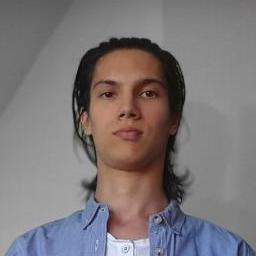}}]{Alexander Balgavy}
is a software engineer, currently building software for the International Space Station at Space Applications Services. He obtained his MSc from Vrije Universiteit Amsterdam, focusing on vulnerability analysis of non-Linux firmware.
\end{IEEEbiography}

\begin{IEEEbiography}[{\includegraphics[width=1in,height=1.25in,clip,keepaspectratio]{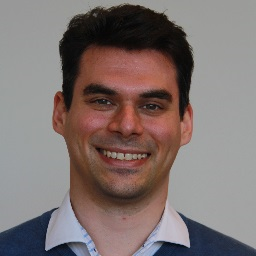}}]{Alexandru Iosup}
is a Full Professor at Vrije Universiteit Amsterdam and leads the AtLarge research group. He has been awarded the Netherlands ICT-Researcher of the Year award, member of the Royal Netherlands Young Academy of Arts and Sciences, and has been knighted by the Romanian Government. 
\end{IEEEbiography}




\end{document}

%% file: sections/0.abstract.tex
Cloud services are critical to society. However, their reliability is poorly understood. Towards solving the problem, we propose a standard repository for cloud uptime data. We populate this repository with the data we collect containing failure reports from users and operators of cloud services, web services, and online games. The multiple vantage points help reduce bias from individual users and operators. We compare our new data to existing failure data from the Failure Trace Archive and the Google cluster trace.

We analyze the MTBF and MTTR, time patterns, failure severity, user-reported symptoms, and operator-reported symptoms of failures in the data we collect. We observe that high-level user facing services fail less often than low-level infrastructure services, likely due to them using fault-tolerance techniques. We use simulation-based experiments to demonstrate the impact of different failure traces on the performance of checkpointing and retry mechanisms.

We release the data, and the analysis and simulation tools, as open-source artifacts available at \url{https://github.com/atlarge-research/cloud-uptime-archive}.

%% file: sections/1.intro.tex
\section{Introduction}\label{sec:intro}

Cloud services are critical to societal processes such as governance, healthcare, finance, science, socialization, and entertainment. However, we have little public data about the reliability of these services. In contrast, our community can access much public reliability on hardware failures~\cite{DBLP:journals/tos/ManeasMES21, DBLP:conf/hotos/HochschildTMGRC21}, failures in datacenter (internal) workloads~\cite{DBLP:conf/dsn/RosaCB15, DBLP:conf/icdcs/El-SayedZS17}, software bug reports~\cite{DBLP:conf/cloud/GunawiHLPDAELLM14}, and other paradigms such as HPC and networks of desktops~\cite{DBLP:journals/jpdc/JavadiKIE13}. To address this problem, we propose collecting \textit{service} failure data from different kinds of services and different vantage points and, in this work, we collect, analyze, and release the data and our tooling.

Cloud services, which are compute services accessed over the Internet, are now embedded into every societal process and have immense impact on society. This impact is similar to that cars had in the 20th century, when they democratized transport. However, cloud services have no standard reliability evaluation, unlike cars, which have institutions like the NHTSA~\cite{nhtsa} and the NTSB~\cite{ntsb} to evaluate their reliability. This makes it difficult for users to evaluate and learn lessons from critical failures of cloud services. A service failure can cause commercial operations to halt~\cite{atlassian22failure}. Cloud-service failures can also have a social impact, such as when students cannot attend classes~\cite{zoom20failure}, public transport is stalled~\cite{ns22failure}, and emergency services stop being available~\cite{kpn19failure}.

The community has proposed several fault-tolerance mechanisms to overcome cloud failures. However, these mechanisms are usually evaluated against simple failure models~\cite{DBLP:conf/nsdi/PrimoracAB21, DBLP:conf/fast/MohanPC21, maurya2024datastates}, or failure traces collected from a single system~\cite{DBLP:journals/tpds/AndreadisMBI22, DBLP:journals/tpds/KadupitiyaJS22}. Often, the traces used are not public~\cite{DBLP:conf/nsdi/BrookerCP20, DBLP:conf/dsn/GuptaTJRM15}. The limited evaluation makes it difficult for researchers and practitioners to evaluate fault tolerance mechanisms comparatively. 


\begin{figure}[t]
    \centering
    \begin{subfigure}[t]{\linewidth}
    \centering
    \includegraphics[width=\linewidth]{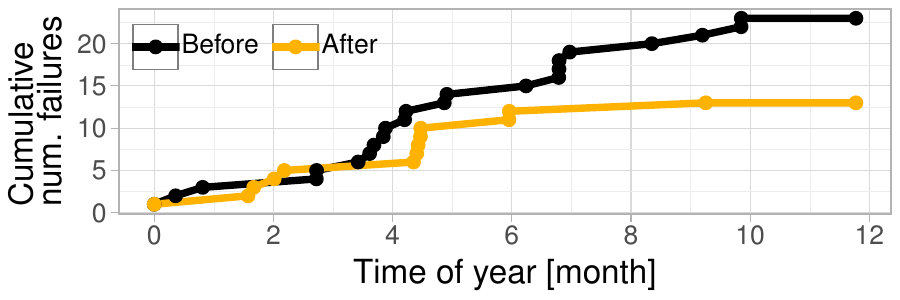}
    \end{subfigure}

    \begin{subfigure}[t]{\linewidth}
    \centering
    \includegraphics[width=\linewidth]{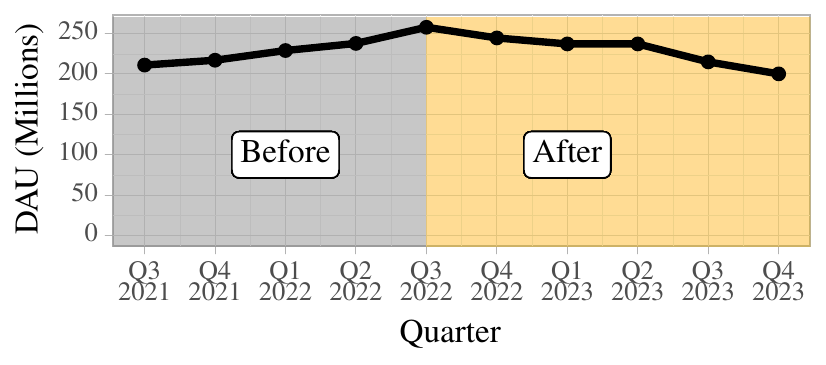}
    \end{subfigure}
    
    \caption{X/Twitter failures and daily active users (DAU) before and after it was taken private. DAU data estimated by Statista~\cite{twitter_dau_statista, twitter_dau_backlinko} and Sensor Tower~\cite{twitter_dau_sensortower} via NBC~\cite{twitter_dau_nbc}.}
    \label{fig:twitter_private}
\end{figure}

We propose the Cloud Uptime Archive (CUA) to remedy the problem. The archive consists of availability traces from different cloud services, and tools to collect and analyze them. The trace archive approach has been successfully used multiple times already to the community's benefit. Example of trace archives include parallel workloads archive~\cite{DBLP:journals/jpdc/FeitelsonTK14}, grid workloads archive~\cite{DBLP:journals/fgcs/IosupLJADWE08}, failure trace archive~\cite{DBLP:journals/jpdc/JavadiKIE13}, cloud software bugs~\cite{DBLP:conf/cloud/GunawiHLPDAELLM14}, and workflow trace archive~\cite{DBLP:journals/tpds/VersluisMTHPDI20}. Closest to this work, the Failure Trace Archive was an early effort in collecting and analyzing computer systems failure traces, focusing on hardware failure traces and workstation availability data. In contrast, the data we collect is about cloud services. Cloud services have multiple software layers between the user and the hardware, making service failure qualitatively and quantitatively different from hardware failure.

As an example of the data we collect and the analysis it enables, we analyze failures of the social network service X/Twitter, before and after it was taken private. \Cref{fig:twitter_private} depicts Twitter's failures collected from the crowdsourced failure reporting website Outage.Report. For the year up to the date Twitter was taken private (28 October 2021 - 2022), Twitter experienced 23 failures with 1,084 user-reports. In the year starting from the day it was taken private (29 October 2022 - 2023), Twitter experienced 13 failures with 199 user-reports. The low number of failures indicates that a large web service can keep operating even after a massive reduction in headcount. However, failures alone do not portray the full picture. We combine failure and user-count data (also depicted in \Cref{fig:twitter_private}) and analyze this further in \Cref{sec:anatomyofafailure}.

Existing work~\cite{DBLP:conf/cloud/PotharajuJ13, DBLP:conf/eurosys/ChenXMKGSCGFWZG24} studies failures reported by their cloud operator. That is not enough, as failure reporting is a subjective process, and the human and economic incentives involved mean that just studying one source of failure information can distort our view of failures. The distortion of focusing on just one source for experiment traces has happened before with workflow traces~\cite{DBLP:conf/usenix/AmvrosiadisPGGB18}. Therefore, to provide a different viewpoint, we collect failure data from multiple aggregators of user-reported failures. For yet another vantage point, we also collect active user counts of online services when possible.

Collecting and characterizing failure reports of cloud services is challenging. First, the data collection is hampered by the lack of persistence in data published by crowdsourced aggregators; most of the data remains available for less than a day. Second, analysis of provided-reported failure data is hampered by inconsistent reporting in natural language. Third, data from different sources comes in different forms and is not directly available as failures. Extracting failures from user reports and active user counts is challenging.

Towards understanding the reliability of cloud services and standardizing the evaluation of fault tolerance mechanisms, we make the following contributions:

\begin{enumerate}
    \item We collect long-term failure data from different types of cloud services, Web, Infrastructure Cloud, and Online Games. Apart from operators' status pages, we collect data from unique vantage points such as crowdsourced failure reporting websites and player counts for online games. We describe the nature of failures, as reported by operators and users, in \Cref{sec:anatomyofafailure}, and compare failures from different sources to validate the dataset in \Cref{sec:validation}.
    
    \item We analyze the MTBF and MTTR, time patterns, and failure severity using statistical techniques in Sections~\ref{sec:tbf}~to\ref{sec:length_impact}. We compare the results of our statistically analysis to existing datasets such as the FTA~\cite{DBLP:journals/jpdc/JavadiKIE13} and Google cluster traces~\cite{DBLP:conf/eurosys/VermaPKOTW15}. 
    
    \item We manually analyze the natural language reports of a one-year-long subset of data, and the rest using LLMs, to identify failure symptoms and root causes in \Cref{sec:causes}.
    
    \item We use traces from our dataset to demonstrate how it can be used to evaluate fault tolerance mechanisms. We evaluate variants of fault-tolerance mechanisms such as checkpointing and retries using simulation in Sections \ref{sec:checkpoint}~and~\ref{sec:retries}.
    
    \item We release the dataset and the tools we use to collect, extract, clean, and analyze the data as FAIR Open Science artifacts. \textbf{The dataset covers 27~data sources} and covers popular services such as infrastructure clouds (AWS, GCP, Azure), infrastructure services (Github, Atlassian), social media (Instagram, Twitter), and online games (Minecraft, Runescape). The dataset and tools are available at \url{https://github.com/atlarge-research/cloud-uptime-archive}. The dataset is also available on Zenodo at \url{https://doi.org/10.5281/zenodo.14712441}.
\end{enumerate}



\section{Anatomy of a Failure} \label{sec:anatomyofafailure}

\begin{figure}[t]
    \centering
    \includegraphics[width=\linewidth]{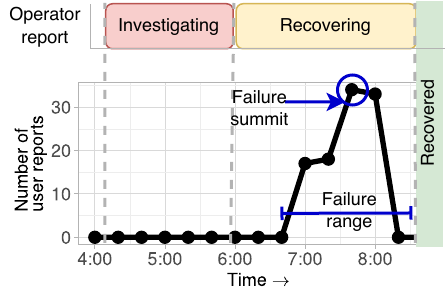}
    \caption{User and operator reports of the Github failure on July 13, 2020. The failure range and failure summit are highlighted.}
    \label{fig:github_13jul20}
\end{figure}

We define failure as when a cloud service cannot provide all or some of its functionality to all or some of its users. A system's true state is only visible to the service operators, and for sufficiently complex systems, even they might not have full visibility~\cite{DBLP:conf/hotos/HuangGZLDCY17}. However, different actors, such as operators, users, and metrics collection systems, can measure the system's state and, hence, the failure state from their own perspective. These different measurement apparatuses are the \textit{vantage points} from which we can analyze a system's failure performance. We describe two examples to demonstrate why the different vantage points are necessary.

For each failure, users report their experience of the failure to a third-party reporting service like DownDetector or Outage.Report. An exemplary sample of such reports for a failure experienced by the service Github is depicted in \Cref{fig:github_13jul20}. We identify all adjacent timestamps with non-zero reports as a \textit{failure range}. The maximum number of reports for a single timestamp in that range is the \textit{failure summit}. The operators also label the failure with different statuses at different times, as depicted in \Cref{fig:github_13jul20}.

\noindent\textit{Short-term dynamics (Github example):} \Cref{fig:github_13jul20} depicts the user reports of a failure experienced by the popular code-hosting service on July 13, 2020. The figure depicts user reports and the status reported by the operator. Github reported this incident as a major failure~\cite{github_13jul20_failure}, and it was covered by news media~\cite{github_13jul20_register}. The operator noticed the failure and started investigating at 4:06 AM UTC. However, no users reported the failures as it was early in Europe and late evening in the USA, where most Github users are from. The operator identified the cause of failure at 5:53 AM UTC and initiated recovery. Users started using the service in the morning and reported the failure to the third-party website. The number of reports increased as the morning progressed. As the operator reached the end of the recovery process, the number of user reports also dropped.

We observe a time lag between when the operator reported the failure and when users observed it. This time gap allows operators to recover from failure without the users even noticing. Systems developers can use the two different failure traces, from operators and users, to evaluate their fault-tolerance mechanisms.




\noindent\textit{Long-term dynamics (X/Twitter example):} Twitter's status page was taken down since twitter was taken private. The exact uptime has not been officially reported or known since. Since being taken private, up to 75\% of Twitter's staff were let go~\cite{twitter_staff_cuts}. Whether the service availability has decreased, as an expected consequence of the layoff, is unknown. We perform a preliminary investigation towards answering this. We compare Twitter's uptime from when it was taken private to the year after in \Cref{fig:twitter_private}. For the year up to the date Twitter was taken private (28 October 2021 - 2022), Twitter experienced 23 failures with 1,084 user-reports. In the year starting from the day it was taken private (29 October 2022 - 2023), Twitter experienced 13 failures with 199 user-reports. The low number of failures indicates that a large web service can keep operating even after a massive reduction in headcount.

We also investigate if the low number of failures is due to less people using twitter. We infer daily active user data (DAU) by combining the DAU data up to 2022 from Statista~\cite{twitter_dau_statista, twitter_dau_backlinko} with the user decrease data for the subsequent year from Sensor Tower~\cite{twitter_dau_sensortower} via NBC~\cite{twitter_dau_nbc}. The DAU data indicates that while Twitter's DAU has decreased, it has not decreased to the extent failures have decreased. The reduction in failures without a significant reduction in DAU indicates that the headcount reduction at Twitter has not had a significant impact on service reliability.

%% file: sections/2.sys_model.tex
\section{Method for Collection and Interpretation of Cloud Uptime Data}\label{sec:method}

\begin{table*}[t]
\centering
\caption{Summary of collected datasets.}
\label{tab:dataset_summary}

\begin{tabular}{@{}lllll@{}}
\toprule
ID & Dataset Name                                                             & Time Period                                                                                                       & Services Tracked                                                                                                                                                         & Fields                                                                           \\ \midrule
WU & \begin{tabular}[c]{@{}l@{}}Web User Reports\\ (Outage Report)\end{tabular} & April 2019 - September 2024                                                                                           & \begin{tabular}[c]{@{}l@{}}Apple Servers, Facebook, FB Messenger, \\ GitHub, Gmail, Instagram, Netflix, Snapchat, \\ Skype, Twitter, WhatsApp, YouTube\end{tabular} & \begin{tabular}[c]{@{}l@{}}Time, number of reports, \\ reason\end{tabular}       \\
CU & \begin{tabular}[c]{@{}l@{}}Cloud User Reports\\ (Downdetector)\end{tabular}  & January 2018 - December 2020                                                                                      & Google Cloud, AWS, Microsoft Azure                                                                                                                                       & Time, number of reports                                                          \\
WO/CO  & \begin{tabular}[c]{@{}l@{}}Web/Cloud \\ Operator Reports\end{tabular}           & April 2019 - September 2024                                                                                           & \begin{tabular}[c]{@{}l@{}}Google Cloud, AWS, Azure, Slack\\ Github, Atlassian services, Discord\end{tabular}                                                                         & \begin{tabular}[c]{@{}l@{}}Time period, affected \\ services, reason\end{tabular} \\
G  & \begin{tabular}[c]{@{}l@{}}Online Game\\Player Count\end{tabular}              & \begin{tabular}[c]{@{}l@{}}August 2020 - May 2021 (Minecraft)\\ August 2007 - March 2015 (Runescape)\end{tabular} & \begin{tabular}[c]{@{}l@{}}Runescape, Minecraft (Hypixel, The Hive\\ Cubecraft, Minehut)\end{tabular}                                                                                & \begin{tabular}[c]{@{}l@{}}Time, number of \\ online users\end{tabular}                                                      \\ \bottomrule
\end{tabular}
\end{table*}

We aim to collect longitudinal availability data for the main ICT services for 21st-dentury economies and societies.
We collect availability data for different cloud services from different vantage points: service operator status pages, user-reports to failure reporting websites like Downdetector, and user counters. The different vantage points each have their biases. By collecting data from all three of them, we expect to get a fuller picture. 

We collect data from different services intermittently over a 17 year period (2007-2024). We collect data from web, cloud, and online gaming services. These categories are representative of what our society finds most lucrative and popular.

We extract failure ranges and peaks from the user reports. We describe the process in \Cref{sec:method:extraction}. We extract failure severity based on number of affected services and number of user reports in \Cref{sec:method:severity}.



\subsection{Data Collection}
\label{sec:method:collection}

We use a web crawler which retrieves raw data from different sources. The different sources are summarized in \Cref{tab:dataset_summary}. We collect data every hour for some sources and every 6 hours for others, based on their update frequency and throttling. For example, we collect data every hour from outage report even though the failure cause data is reset every 20 minutes, to avoid throttling. We then parse the raw data, which is in HTML or JSON format. We had to design and implement custom parsers for each data source to extract the required information.

One obstacle we faced while collecting user reports of cloud failures was that failure reporting websites like Downdetector blocked and throttled our data collection efforts. We were able to obtain historical data using distributed household proxies. However, even these effort were successfully blocked. Hence, that dataset ends in December 2020.

\subsection{Failure Extraction}
\label{sec:method:extraction}

Some data sources, such as operator status pages, provide a clear failure period with a start time and an end time. But other crowdsourced sources only provide number of user reports of failures (Outage Report and Downdetector) or number of online users (Runescape and Minecraft). To extract failure data from these crowdsourced sources, we detrend the data and then use peak extraction. Our hypothesis is that failures are period with with a sudden increase in the number of user reports of failures or a sudden decrease in the number of user.

For preprocessing, we check whether there is a seasonality, meaning a recurring pattern with a fixed period. If such a pattern exists in the data, it means that the fluctuations caused by that pattern are not anomalous. To facilitate finding the points which fall outside of this regular pattern, it is useful to remove the seasonal component from the data. The data may contain a long-term trend, which we also remove.

To extract and remove these two components from the data, we use STL~\cite{cleveland1990stl}. 
STL decomposes the time series into the trend component, the seasonal component, and the remainder. This remainder is useful for the identification of anomalies, as it contains the deviations from the regular patterns. We chose STL because of it supports arbitrary sampling frequencies and because its simplicity allows for fast computation; something which is rather important given the high-frequency data which, in the case of Runescape, has over 2 million data points.

\begin{figure}[t]
    \centering
    \includegraphics[width=\linewidth]{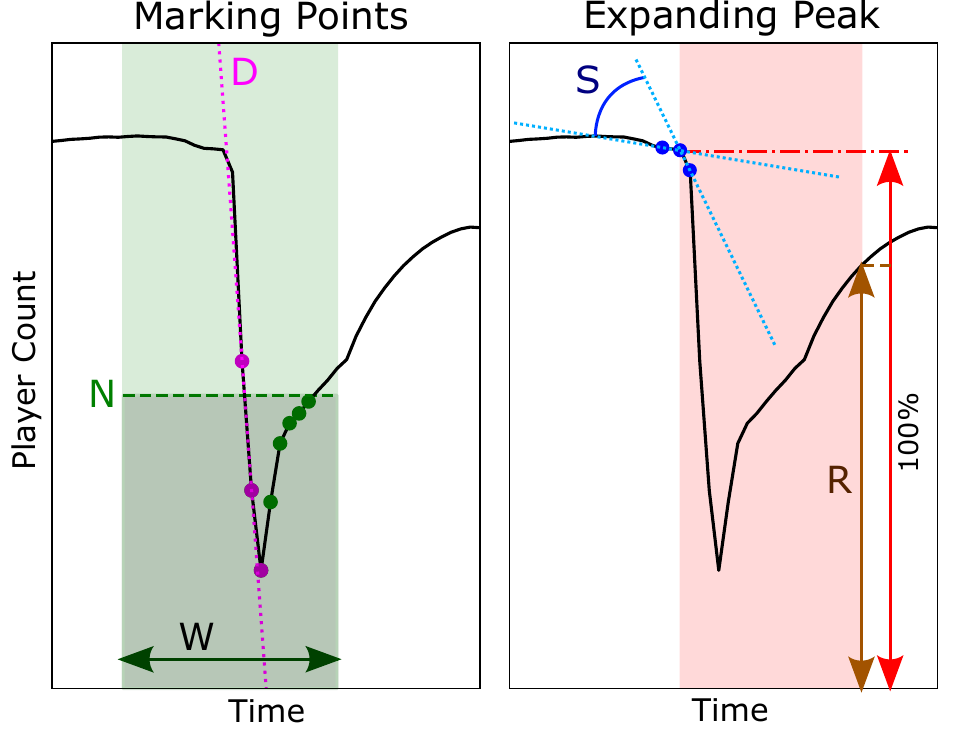}
    \caption{Peak detection algorithms.}
    \label{fig:peak-detection}
\end{figure}

We employ two methods for peak detection. One to identify peak location and another to identify the duration of the peak. Both methods are depicted in \Cref{fig:peak-detection}. This combination of location and duration correspond to a failure.

The first method (illustrated in the left subfigure) calculates the change in the number of user reports or users over a rolling window (W), operating on the deseasonalized dataset. When a data point is (N) standard deviations below the mean of the window, it is marked as an anomalous point. The magenta line (D) represents the first-order difference threshold for marking points.

To identify failure duration, the anomalous points are expanded backwards and forwards in time. This is illustrated in the right subfigure. For each point marked as anomalous, its preceding points are evaluated one by one until a sharp drop in the first-order difference (S) on the raw value is observed. The peak is considered to have ended when a certain fraction of players (R) have been regained. The red arrow shows 100\% of the original player count for reference.

\subsection{Natural Language Report Data Extraction}
\label{sec:method:llm}

Cloud services report failure information such as the root cause and affected services in natural language. Extracting key information from failure reports is time- and labor-consuming because the reports have idiosyncratic formatting and are written by different people. 

We use two different approaches for this. First, we apply large language models (LLMs) to automatically extract high-confidence fields from failure reports. Second, we use manual human analysis for low-confidence fields to extract information from a one-year subset of the data.  

To determine the confidence of a field, we first split the data into test and validation sets. The validation set comprises of 1-3\% of the total reports for each provider. We manually extract the fields we need such as the service name, service category, location, user symptoms, and root cause for each report in the validation set. Then, we use LLMs to extract the same information for the validation set. We compare the results from the LLMs to our manual classification results. We consider high-confidence fields as those for which the LLMs have over 70\% accuracy.

To prevent bias from a single LLM, we use three LLMs and have them vote on the correct answer. For the vast majority of failure reports, all three LLMs unanimously agree on the answers. The LLMs we use are: ChatGPT-4, Claude, and Gemini Pro. All of them were accessed through their respective APIs.

We use the method we describe above to study the failure reports from three cloud operators: AWS, Azure, and Google Cloud. We present the results of the analysis in \Cref{sec:causes}.

\subsection{Interpreting Severity}
\label{sec:method:interpretation}\label{sec:method:severity}

The datasets we collected have different descriptions and interpretations of failure. But a unified representation is better for using the data to evaluate fault tolerance techniques. An important part of this is quantifying the severity of a failure. A failure's severity can be based on the number of people affected, the qualitative way they were affected, the number of services affected, and the duration. 

We focus on unifying three different failure severity metrics into a unified format. One metric is the severity of the failure defined by the cloud operator. Cloud operators define this based on the number of people affected, the impact of the failure, and other considerations.

A second metric is the number of user reports of failure. We first extract failures from user report data using the peak detection technique from \Cref{sec:method:extraction}. We then identify the 5th and 95th percentile of user reports for all extracted failures. We set any value above the 95th percentile as failure severity 1 and any value below the 5th percentile as failure severity 0. The values in the middle are interpreted to be between 0 and 1 using min-max scaling.

The third metric we use is the number of online users. We use the same method that we use for the user reports, except the direction of severity is reversed. The periods with the least amount of online users are failure severity 1 and the ones with the highest amount of users are severity 0, meaning no failure.

A convenient result of interpreting failure severity as lying between 0 and 1 is that we can use these as the probability of a service failing to respond in experimental evaluation. We use the failure severity value as the fraction of machines that failed in \Cref{sec:checkpoint}. We use it as the fraction of network requests that failed in a given time period in \Cref{sec:retries}.

\begin{figure}
    \centering
    \includegraphics[width=0.9\linewidth]{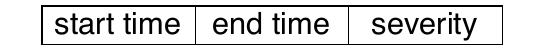}
    \caption{CUA trace format.}
    \label{fig:cua_format}
\end{figure}

\subsection{Cloud Uptime Archive Data Format}

We release both the raw failure data we collect and failure traces in a normalized format. The normalized traces are made for easy consumption by a real-world experiment or simulation. We describe the normalized trace format here. If users want granular control, they can always fall back to the raw data.

Each file in the dataset represents the failure trace for a single service, from a single vantage point. For example, Github (service) failures reported by users (vantage point). This information is incorporated into the filename, e.g., \textit{cua\_github\_outagereport.csv}. Each file is in CSV format and contains multiple lines. Each line represents a failure. In the normalized format, failures do not overlap. However, in some raw data sources where each service provided by an operator is tracked independently, failures can overlap. We combine them into a single failure during normalization.

\Cref{fig:cua_format} depicts the format of each line in a trace file. The start time and end time represent the start and end of the failure. These times are not normalized to start at 0. The first failure does not necessarily occur at the start of the experiment. Users should normalize the timestamps in the trace to their desired start time. The final field is normalized severity as we described in \Cref{sec:method:severity}. It has a value between 0~and~1. 0~represents no failure and 1~represents total failure. This can be used as the probability that a machine or set of machines fail, as we do in \Cref{sec:checkpoint}. It can also be used as the probability that a request fails, as we do in \Cref{sec:retries}. The severity can also have other interpretations depending on the user's experiment.

%% file: sections/3.selection_and_validation.tex
\begin{figure*}[t]
    \centering
     \begin{subfigure}[b]{\textwidth}
         \centering
         \includegraphics[width=\textwidth]{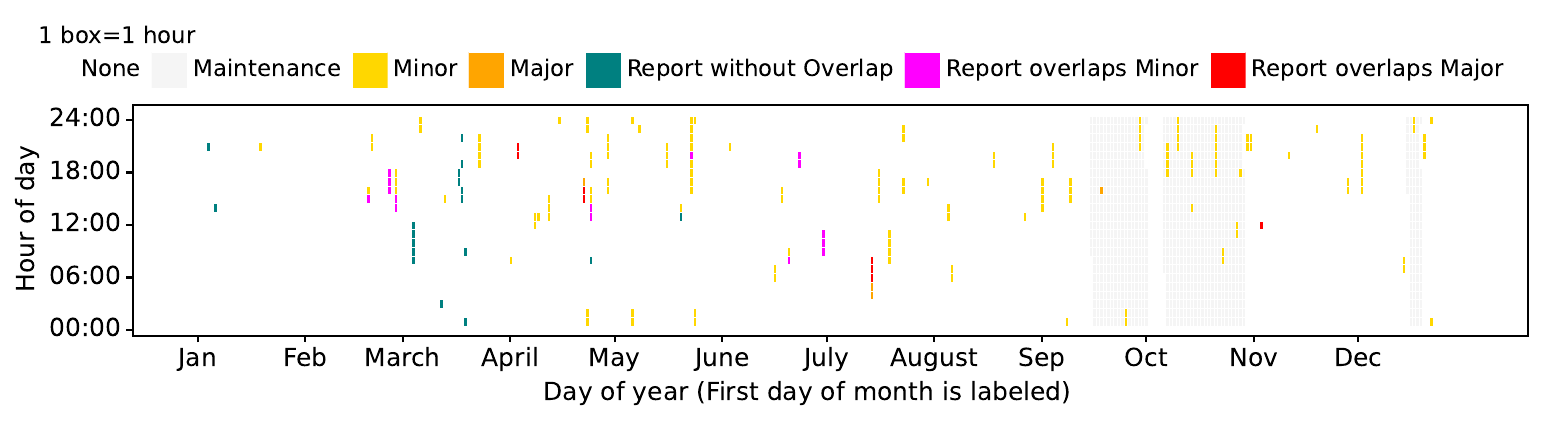}
         \caption{Github}
         \label{fig:validation:github}
     \end{subfigure}
     \hfill
     \begin{subfigure}[b]{\textwidth}
         \centering
         \includegraphics[width=\textwidth]{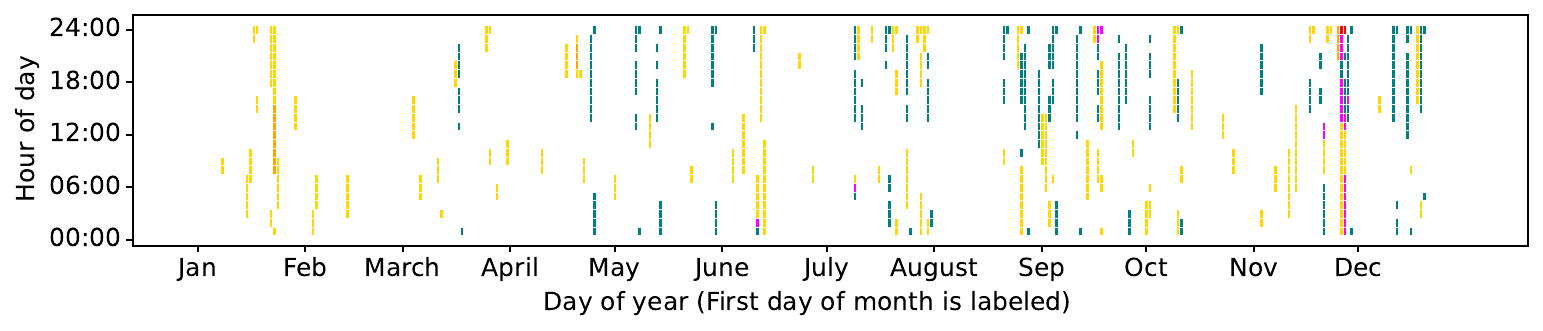}
         \caption{Amazon Web Services}
         \label{fig:validation:aws}
     \end{subfigure}
     
     \caption{Failure ranges detected from user- and operator-reported failures. Operators report maintenance, minor, and major failures. Users report incidents, which we process for this work and check for overlaps with operator reports.}
     \label{fig:validation}
\end{figure*}

\section{Validation by Comparing Data from Different Sources}
\label{sec:validation}

To assess the usefulness of user reports in identifying failures, ideally, we would compare them to the ground truth of failures for all services we monitor. However, this is not possible, as the operators rarely publicly report all the failures their cloud services experienced. Instead, we take as ground truth the set of official updates about failures currently reported on each cloud service’s status page. We have such information from status pages of several operators such as Github, Atlassian, AWS, GCP, Azure, and more. 
This allows us to show evidence of whether the failures reported by users are similar to the failures the operator self-reports. Furthermore, this gives insight into whether the metrics cloud services are tracking are what users experience as failures or if users experience something different. Our main findings are:

\begin{description}
    \defmainfinding{mf:github-majormatch}{There is overlap between the \textit{major} failures on the operator status page and those reported by users.}
    
    \defmainfinding{mf:github-does-not-match}{There is little overlap between \textit{all} the failures self-reported by the operator and crowd-reported by users.}
\end{description}

Figure~\ref{fig:validation} depicts Github and AWS failures as (officially) self-reported by the operator and as reported by users to Outage Report. The horizontal axis represents the days of the year. The first day of each month is labeled. The vertical axis represents the time of the day. A single box represents one hour of one day. There are 5~major issues officially reported by Github and 4~major issues reported by AWS. 4~out of the 5~major Github issues overlap user reports to Outage Report~(\refmainfinding{mf:github-majormatch}). 2~out of the 4~major issues AWS issues overlap issues reported by Outage Report. We further find 8~and~10 official minor failures that overlap with user reports for Github and AWS, respectively. This gives evidence the overlap cannot occur randomly, so \textit{user-reported failures are consistent with the ground truth}.

We also observe that many official minor reports have \textit{no} overlap~(\refmainfinding{mf:github-does-not-match}). This suggests little overlap between what users and GitHub perceive as important enough to report. The difference between the sets of official and user reports stems from the different kinds of failures they track. Minor events reported by Github range from degraded performance to problems with forking and accessing the website. In contrast, the most popular reason for events reported to Outage Report by far is ``Website Down'', which accounts for about two-thirds of the reports for Github. In line with similar conjectures about software-developing companies~\cite{DBLP:journals/cacm/BouwersVD12}, we conjecture the official reports correspond to failures tracked by Github's {\it internal} metrics, whereas user-reports correspond to events with high visibility for the {\it (external) users}.

The difference between officially reported and user-reported failures suggests a gap in the definition of failures that are to be reported. On the one hand, as researchers and expert users, we want detailed information when a service we use is not functioning. On the other hand, a service cannot publish thousands of small events as failures because this could fatigue the users and be misinterpreted as pervasive unreliability. 

\begin{figure*}[t]
    \centering
    \begin{subfigure}[t]{0.49\textwidth}
        \centering
        \includegraphics[width=\linewidth]{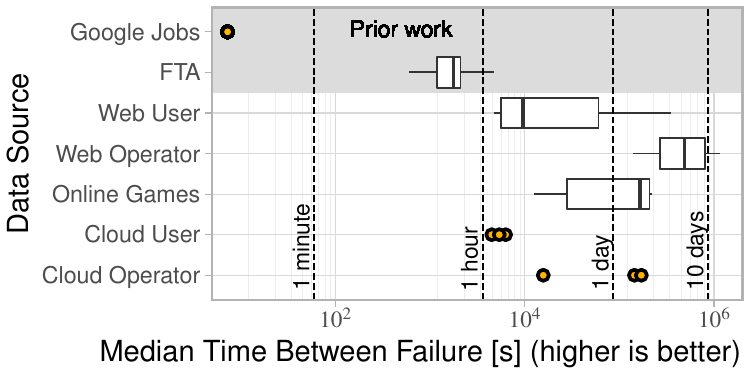}
        \caption{Distribution of median time between failure of traces across categories.}
        \label{fig:mtbf-median}
    \end{subfigure}
    \begin{subfigure}[t]{0.49\textwidth}
        \centering
        \includegraphics[width=\linewidth]{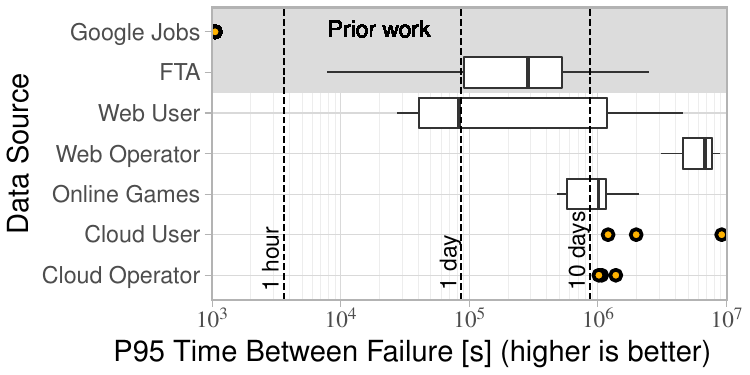}
        \caption{Distribution of P95 time between failure of traces across categories.}
        \label{fig:mtbf-tail}
    \end{subfigure}

    \begin{subfigure}[t]{0.49\textwidth}
        \centering
        \includegraphics[width=\linewidth]{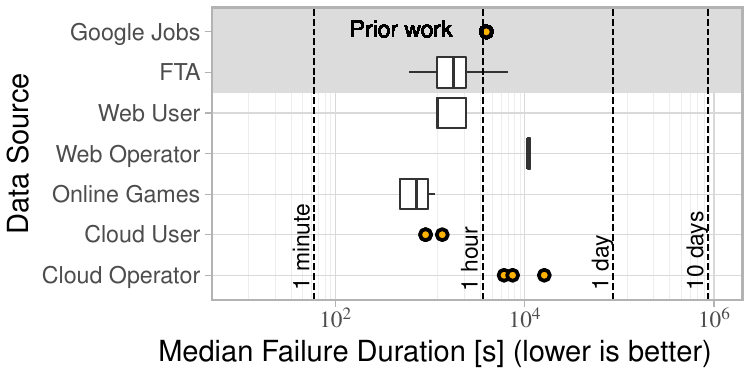}
        \caption{Distribution of median time to repair of traces across categories.}
        \label{fig:mttr-median}
    \end{subfigure}
    \begin{subfigure}[t]{0.49\textwidth}
        \centering
        \includegraphics[width=\linewidth]{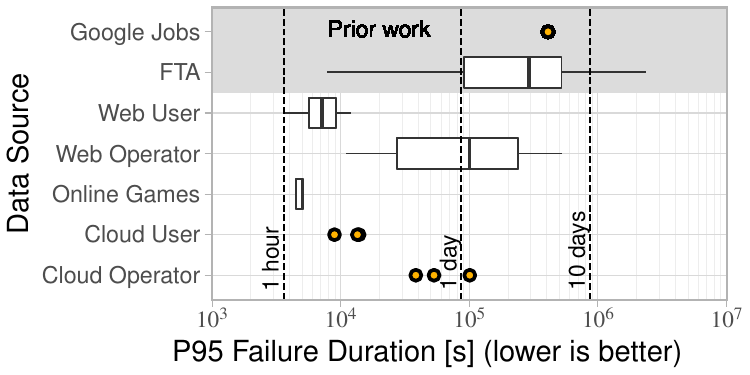}
        \caption{Distribution of P95 time to repair of traces across categories.}
        \label{fig:mttr-tail}
    \end{subfigure}

    \caption{Statistical properties of the time between failures (time between the end of a failure range and the start of the next one) and failure duration (time between the start of a failure range and its end). Entries taken from prior work are marked with a dark background. Logarithmic horizontal axis, with visual aids for minute, hour, day, and 10 days.}
     \label{fig:mtbf}\label{fig:mttr}
\end{figure*}

Our findings in this section motivate the need to revisit current failure-reporting methodologies, to make them identify and classify failures consistently with what the users experience. 
We recommend reporting both global failures and fine-grained, detailed failures per operational instance (shard) of each service; for example, the Salesforce status-page~\cite{salesforce_statuspage} offers already such reporting.

%% file: sections/4.mtbf_and_mttr.tex

\section{Time Between Failure and To Recovery}\label{sec:tbf}

\begin{figure*}[t!]
    \centering
    \includegraphics[width=0.99\textwidth]{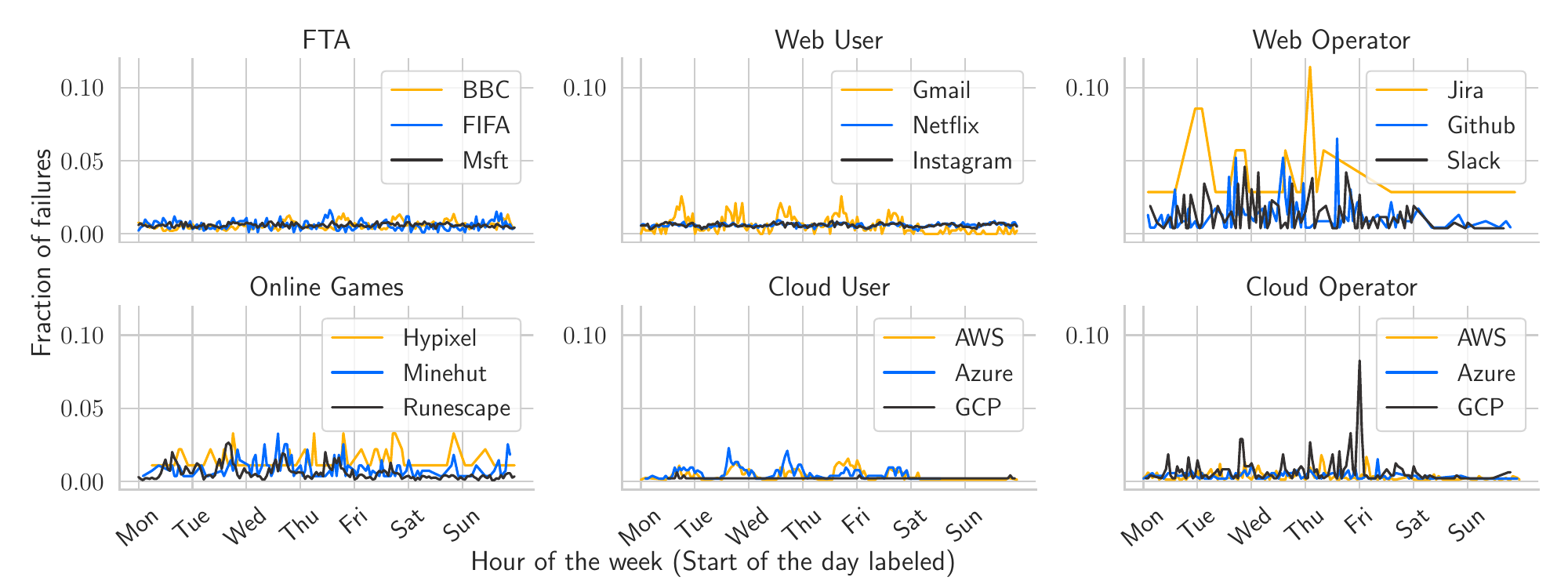}
    \caption{Weekly trend of failure aggregated per hour. Breaks on the horizontal axis at the start of each day.}
    \label{fig:weekly_trend}
\end{figure*}

We investigate the difference in cloud service reliability across different datasets and layers of abstraction and make the following observations:

\begin{description}
    \defmainfinding{mf:current-reliable}{Present web and cloud services are more reliable than past web services.}
    \defmainfinding{mf:cloud-less-reliable}{Cloud services are less reliable than web services, indicating that web services, built on top of cloud services, use additional fault-tolerance mechanisms.}
    \defmainfinding{mf:user-frequent}{User-reported failures occur more frequently than operator-reported ones but are shorter.}
    \defmainfinding{mf:slow-recovery}{The median recovery time of web and cloud hasn't improved improved in the last 20~years, but the tail recovery time has improved siginificantly.}
\end{description}

We compare the data from the 5~new datasets we introduce with the service reliability data from the Failure Trace Archive (FTA)~\cite{DBLP:journals/jpdc/JavadiKIE13} and the reliability of datacenter internal workloads from Google~\cite{DBLP:conf/eurosys/VermaPKOTW15}. We use the Median Time Between Failure (MTBF), Tail [P95] Time Between Failure (TTBF), Median Failure Duration (MFD) (We use the term Failure Duration instead of Time to Recovery as in MTTR), and Tail [P95] Failure Duration (TFD) metrics to compare the datasets. These metrics capture properties about both service availability and service recovery which contribute to reliability. We choose P95 for the tail as failures are rare events, which means there are at most 100s of them per service. Using higher values like P99 would only capture the worst couple of events and skew the results.

Comparing the reliability across different levels of abstractions enables us to investigate if services offered at higher abstraction levels (e.g.: Applications) are more reliable than the those offered at lower abstraction levels (e.g.: Infrastructure). Analyzing reliability of both median and tail failures and comparing to the past enables us to investigate where the community has investigated resources to improve reliability.

Figure~\ref{fig:mtbf} depicts boxplots of the MTBF, TTBF, MFD, and TFD of all the traces in each category. The cloud user and operator categories have individual data points instead of a boxplot as those categories only contain 3~traces each (AWS, Azure, and GCP). We 

The median MTBF of FTA is 10x lower than the median MTBF of the Web User category and 100x lower than the Web Operator category in \Cref{fig:mtbf-median}. The median TTBF of FTA at 3 days is 3x longer than the median TTBF of the Web User Category, but over 10x lower than the median TTBF of the Web Operator category in \Cref{fig:mtbf-tail}. This indicates that failure in current services occur much more infrequently than failures in the past (\refmainfinding{mf:current-reliable}).

The MTBF of cloud services is lower than the MTBF of web services, except for user reports for web services (\refmainfinding{mf:cloud-less-reliable}). This implies that web services which want realibility need to use fault tolerance mechanisms of their own to operate in the presence of cloud faults. These mechanisms can include replicas on multiple clouds and failover mechanisms when some cloud services fail.

User-reported failures have a lower MTBF throughout the board for both user- and operator-reported categories (\refmainfinding{mf:user-frequent}). This indicates that operators do not do report failures which don't affect a significant number of users. It could also indicate that the problem is with the users end of the network or the users device itself.

In \Cref{fig:mttr-median}, the median MFD of current web services is either the same (Web User) or 5x higher (Web Operator). The situation is similar for cloud services (\refmainfinding{mf:slow-recovery}). This indicates that the industry has made little progress in improving recovery time of the median failure in the last 20~years. However, the situation is much better for TFD. The median TFD of both web and cloud services is 4-10x better.

%% file: sections/5.time_pattern.tex

\section{Patterns over Time}\label{sec:patterns}

We investigate failure occurrences at different times during the week and make the following observations:

\begin{description}
    \defmainfinding{mf:weekdays}{Current web and cloud services fail more on weekdays compared to the past when the failure was evenly distributed across all the days.}
    \defmainfinding{mf:evening}{Both web and cloud services fail more in the evening (US East) / afternoon (US West).}
    \defmainfinding{mf:opmodes}{Web services do not display clear modes of early operation and deprecation like hardware for the duration of data we collect.}
\end{description}

\begin{figure*}[t!]
    \centering
    \includegraphics[width=0.9\textwidth]{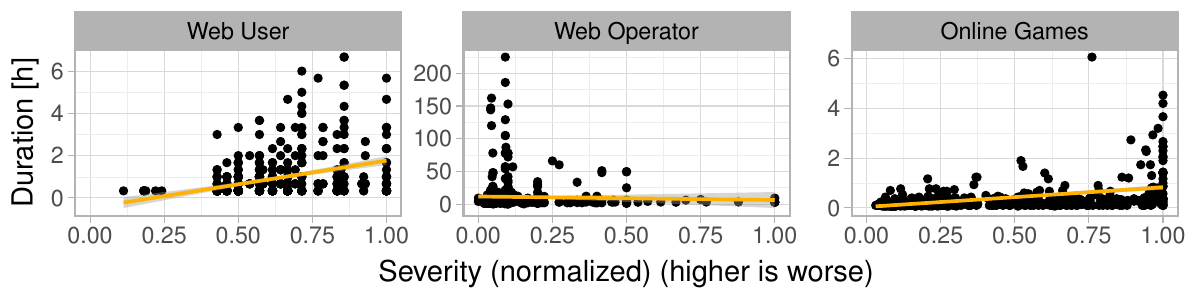}
    \caption{Failure duration compared to number of players for online games.}
    \label{fig:duration_vs_severity}
\end{figure*}

\begin{figure}[t!]
    \centering
    \includegraphics[width=0.7\linewidth]{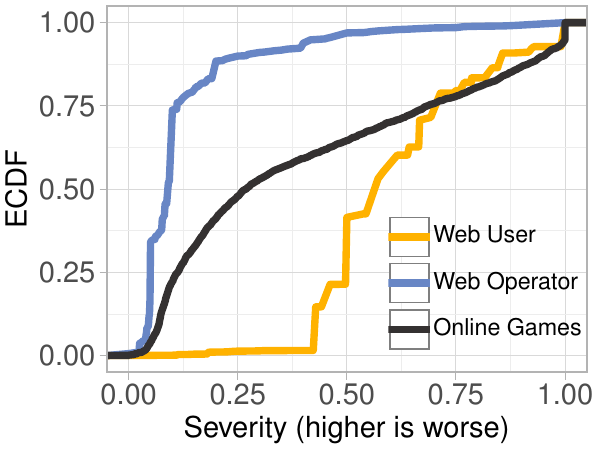}
    \caption{CDF of failure severity, different services.}
    \label{fig:severity}
\end{figure}

Comparing the number of failures across the week for current and past services gives us a picture of the operations of these services. We investigate if services fail more when they are used more, e.g., entertainment services in the evening. We also investigate if services fail less on the weekends, which indicates that the operations processes ensure that new versions are not deployed at the end of the week when engineers do not work.

\Cref{fig:weekly_trend} depicts the failure pattern of service failure in 6~categories across the week. We only depict 3~representative service in each category. The failure aggregated at one hour intervals based on the time of the week they occur. The horizontal axis of the plot denotes the hour of the week, with the breaks representing the start of every day. The vertical axis represents the fraction of total failures that occurred in a specific hour.

We observe that web and cloud services fail more often during the weekdays than during the weekends. This is a contrast to the FTA, where failures are uniformly distributed across the week (\refmainfinding{mf:weekdays}). This indicates that current services have better operations than in the past, which ensure new version are not deployed on or close to the weekends. Better operations means better service version deployment processes, better quality assurance, and other such processes. We conjecture that the failures are caused by new service version deployment and not hardware failures or usage pattern changes, because hardware and users do not change on the short week-level time span. 

We also observe clear spikes in the middle of the week (Tue-Thu) for Web Operator, Cloud User, and Cloud Operator indicating that new versions are deployed in the middle of the week, giving engineers time to tackle problems when they occur.

We also observe that both web and cloud failures are occur in the evenings. Notice that the spikes are very close to the day breaks which indicate midnight. The spike for GCP is particularly visible in the Cloud Operator plot. We believe this is an artifact of our data collection process. We collect data from sensors located on the US East coast. Evening on the US East coast is afternoon on the US West coast. The users of services we investigate are mostly from the USA and this is the time when the largest number of users are online. Large engineering offices of most of the services we investigate are located on the US West coast, which leads us to conjecture that new service versions are deployed when its afternoon in US West.

The online game services hypixel and minehut are more likely to fail at the end of the week, which is different from web and cloud services. This is likely because users play games on the weekends. Runescape does not show this trend and fails more often during weekdays like other services.

Both Schroeder~\cite{DBLP:journals/tdsc/SchroederG10} in 2006, and Amvrosiadis~\cite{DBLP:conf/usenix/AmvrosiadisPGGB18} in 2019 consider trace duration important Schroeder to differentiate between early and late life systems. Amvrosiadis to capture the long-term time dynamics of longer than a week. Our multiple-year long data capture duration means we capture patterns of longer than a week making our traces a realistic and less biased input than only week long traces.

We do not observe an distinct long-term operational modes such as early stage (infancy, 6~months to 2~years) and late stage (obsolescence, last 1-2 years). This is because none of the services that we collect data from have become obsolete during our collection period. The services we investigate are much larger than any single hardware cluster and are composed of multiple services themselves. We cannot differentiate these parts well enough to investigate their operational phases well at this time.




%% file: sections/6.severity.tex
\section{Failure Severity and Impact on Failure Duration}\label{sec:length_impact}
\label{sec:severity}

We investigate failure severity and its relation to failure duration and make the following observations:

\begin{description}
    \defmainfinding{mf:medium-severity}{Operators report most failures to be low severity, whereas most failures are of medium severity going by user reports and user counts.}
    \defmainfinding{mf:severity-corr}{User reports and online player count denominated failures have their duration positively correlated with the severity. Long operator reported failures are of low severity.}
\end{description}

The severity of failures is representative of how many users and how many components of a service a failure effects. We describe how we interpret and normalize severity across datasets in \Cref{sec:method:severity}. We interpret failure severity from user reports based on the number of report; from operator reports based on the number of services and the operators self-reported severity; and from online games based on reduction in player counts.

Understanding failure severity helps engineers evaluate the reliability of a service and plan resource allocation. For example, many low-severity failures indicate that quality assurance and development processes must be improved over the long term. Many high-severity failures indicate that reliability should be the development focus, even to the exclusion of other features.

\Cref{fig:severity} depicts the ECDF of failures for three categories. The horizontal axis is the normalized severity. We describe how we arrive at the normalized severity score in \Cref{sec:method:interpretation}. We observe that web operators report most failure to be not severe. On the other hand, both online games and web users report most failures to be of moderate severity (\refmainfinding{mf:medium-severity}). This indicates that users perception of failure is different from the operators perception. For example, an operator might only be seeing a slight drop in user interaction leading them to classify a failure as low severity. But, the failure might affecting only light users of the service, who, while numerous, might not show up in the metrics as significant because they make few requests each.

\Cref{fig:duration_vs_severity} uses a scatter plot to depict the relation between the severity and duration of failures. For both web users and online games categories, failure duration exhibits a mild positive correlation with severity. This means a more severe failure is more likely to last longer. But, the correlation is not strong, with a Pearson correlation of 0.3 (\refmainfinding{mf:severity-corr}). The web operator category does not exhibit any correlation between failure duration and severity. This is likely because operators report most failures to be of low severity.

%% file: sections/7.causes.tex
\section{Failure Symptoms and Causes}\label{sec:causes}

\begin{figure}[t!]
    \centering
    \includegraphics[width=0.49\textwidth]{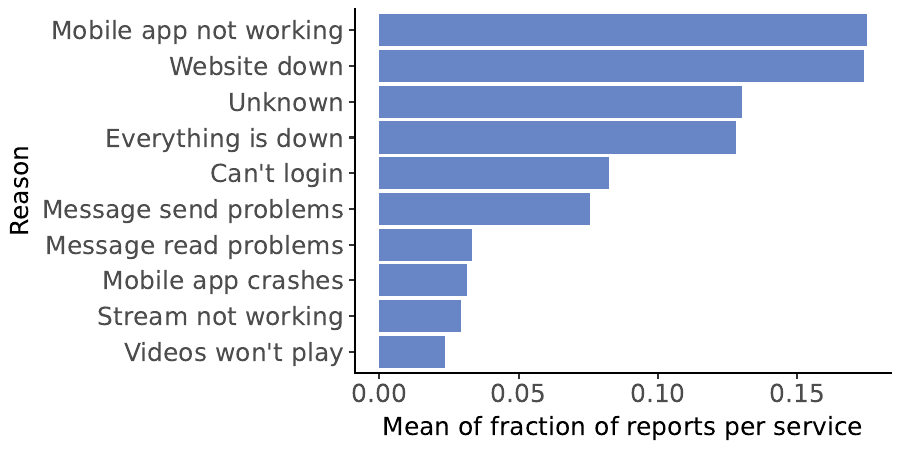}
    \caption{User-reported symptom of failure.}
    \label{fig:web_report_cause}
\end{figure}

\begin{figure}[t!]
    \centering
    \includegraphics[width=0.49\textwidth]{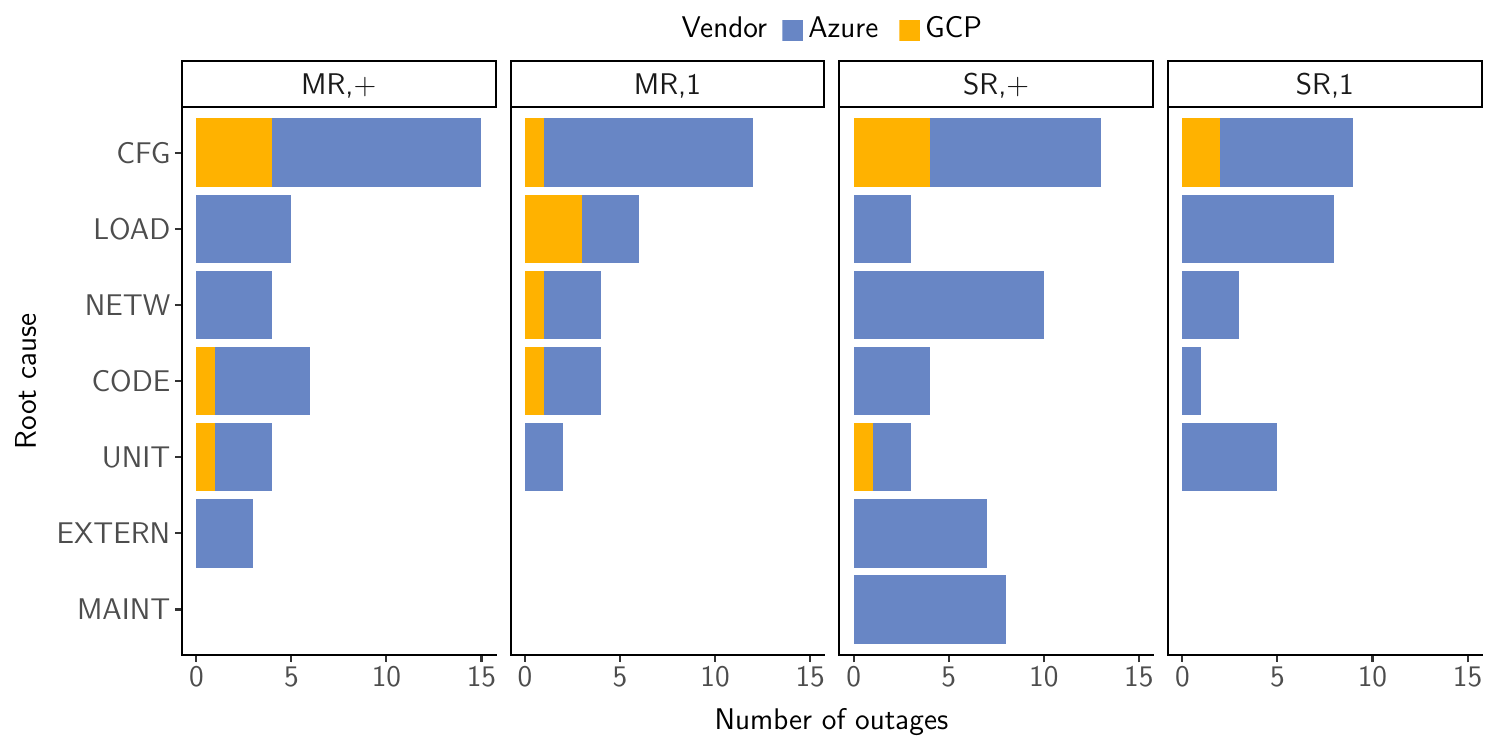}
    \caption{Root cause of failures, along with number of affected regions, for cloud providers.}
    \label{fig:cloud_root_cause}
\end{figure}

\begin{figure}[t!]
    \centering
    \includegraphics[width=0.8\linewidth]{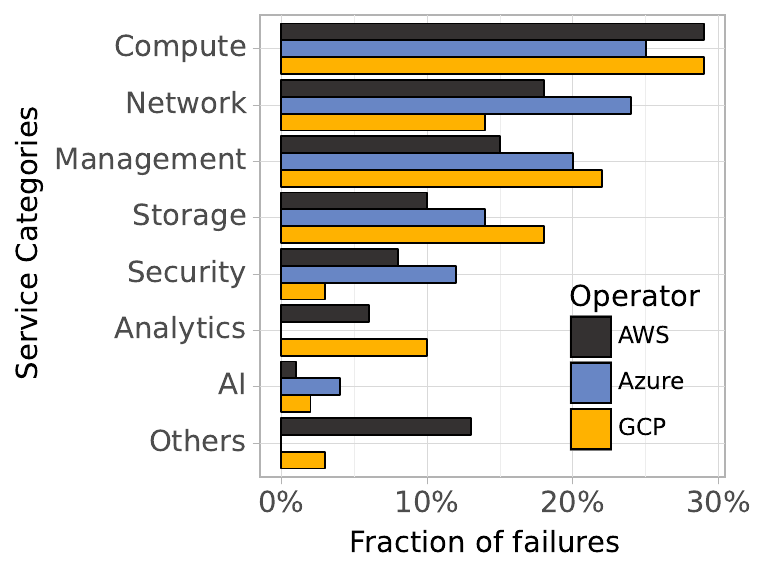}
    \caption{Categories of services affected by failures.}
    \label{fig:cloud-service-cats}
\end{figure}

We investigate the root causes of cloud failures and the symptoms as reported by the user and make the following observations:

\begin{description}
    \defmainfinding{mf:web-mobile}{Most users report a low-information failure symptom and access the service from a mobile phone.}
    \defmainfinding{mf:config-cause}{Most cloud failures are caused by configuration errors, followed by overloads and network issues.}
    \defmainfinding{mf:base-service}{Infrastructure service fail more often than user-facing services.}
\end{description}

Identifying the root cause of failures allows organizations to better allocate infrastructure and developer resources. For example, hardware failures indicate the need to invest in fault-tolerance techniques such as checkpointing and replication. Bugs in the code and configuration indicate the need for better developer tools and quality assurance mechanisms. We analyze natural language failure data from the year 2020 for both user reports and infrastructure-cloud (AWS, Azure, GCP) operator reports of failure.

Outage Report provides users the choice to choose the symptom of failure they experience while reporting. These symptoms are unique to the service such as ``Mobile app not working" for Instagram and "Unable to push" for Github. 

We summarize the reported symptom across all services for the top ten symptoms. To compute this summary, we first compute the fraction of reports of each service which were reported due to a particular symptom. Then, for all symptoms of all services, we compute the mean fraction by adding the fraction of the symptom for all service and dividing by twelve, the number of services. We plot the ten symptoms with the highest mean in Figure~\ref{fig:web_report_cause}. 

We observe that the most popular failure symptom is ``Mobile app not working". Notice that ``Mobile app crashes" is also another reason further down the list. The likely reason for this is that for all services except Apple and Github, the mobile app is a major way people interact with these services. We do not know if the problem is with the mobile app or the cloud service behind the app. In either case, this is a kind of failure that should be further studied by the community. We should investigate if mobile applications present opportunities to deal with failure that were previously unavailable. The second most popular symptom is ``Website down"~(\refmainfinding{mf:web-mobile}). This likely refers to the website not loading at all or being very slow to load that it is unusable. We expected this to be a popular symptom as it is a catchall for all kinds of failures that might crop up in application accessed via the browser. Another such catch all which is further down the list is ``Everything is down." All reports without a reason were categorized as ``Unknown."

\Cref{fig:cloud_root_cause} depicts the root cause given by two cloud operators (Azure, GCP) in their own failure reports. We perform this analysis manually over a one year subset of the data. We analyze the distribution of reasons across root causes and impact levels. We identify seven classes of root causes: UNIT (individual nodes, instances, or clusters, not necessarily hardware), NETW (related to the internal or external network), MAINT (side effects caused by maintenance), LOAD (increased load on the service), EXTERN (external causes, i.e. environmental or third-party), CODE (code errors/bugs), and CFG (configuration errors). We further separate the outages by vendor, indicated by the color of the bar in Figure 3. We do not include AWS, as we do not have sufficient data regarding root causes from AWS (a cause was only specified for 2.1\% of AWS events). Excluded from the plot are outages that did not provide a root cause (a total of 122 events, 45.69\%), a range (5 events, 1.87\%), or the number of affected services (4 events, 1.5\%).

The majority of outages across all levels of impact are caused by configuration errors. For multi-regional outages, the configuration category accounts for the majority by a wide margin. (\refmainfinding{mf:config-cause}) On the other hand, for single-region outages, there are multiple leading causes: apart from configuration errors, outages affecting one service are also caused by increased load and failing instances, and those affecting more than one service are also caused by network errors and maintenance side effects.

We observe that load-related outages tend to happen starting between midday on Wednesday and Thursday evening, with smaller peaks during the night on other days. Most outages caused by misconfiguration happen in the first half of the week, from mid-Monday until midnight on Thursday. This could perhaps be because large code changes are usually introduced during the first few days of the week. The major peaks for these outages generally occur around midnight. The reason for this could be that deployments happen during the night, when it is likely that fewer customers will be using the service; as Langford et al. found, there is a significant difference in traffic during the day and during the night~\cite{DBLP:journals/isem/LangfordLMP12}. It could also be that changes are deployed during the day, but there is a delay before bugs appear.

We identify the service categories that were affected by the failures. We use LLM-enabled analysis of operator failure reports (AWS, Azure, GCP) using the method we describe in \Cref{sec:method:llm}. The results of our analysis are depicted in \Cref{fig:cloud-service-cats}. We classify cloud services into 8~categories. We then use LLMs to classify the failure reports into these categories. We observe that infrastructure services like compute services (e.g., EC2 in AWS), network services (e.g., VPC in AWS), and the management plane fail most often. Services which offer a higher level of abstraction and are more user-facing, such as Analytics and AI fail the least (\refmainfinding{mf:base-service}). This supports our earlier observations in \Cref{sec:tbf} that user-facing Web services have higher inter-failure duration than infrastructure cloud services.

%% file: sections/8.checkpointing.tex
\section{Using failure traces to evaluate the impact of failures and checkpointing on HPC applications}
\label{sec:checkpoint}

We demonstrate how the traces we release in this work can be used to investigate the fault-tolerance properties of a system or an operational technique. This tackles the challenge we highlight in the introduction, where most fault-tolerance papers use synthetic or nonpublic traces making comparison difficult.

We investigate the impact of using a checkpointing approach on an HPC cluster using failure traces and make the following observations: 

\begin{description}
    \defmainfinding{mf:checkpoint1}{The best checkpointing model depends both on the workload executed and on the frequency of failures. Selecting a good checkpointing model can decrease the runtime of a workload by 10-35\% for medium-frequency failure traces.}
    
    \defmainfinding{mf:checkpoint2}{Using checkpointing does not always improve performance and can even reduce it. Selecting an incorrect model increases the average task delay by up to 25\% compared to using no checkpointing when a workload is exposed to low-frequency failure traces.}
    
    \defmainfinding{mf:checkpoint3}{The effect of failures and checkpointing can be much greater on single customers than on overall performance. Selecting the best checkpointing model reduces the average Task Delay by 48\%, while only reducing the overall runtime by 2\% when a workload is exposed to low-frequency failure traces.}
    
    \defmainfinding{mf:checkpoint4}{There is no single checkpointing model that always outperforms all other models. Instead, three of the five checkpointing models used alternate between performing the best depending on the workload and failure trace.}
\end{description}

\subsection{Checkpointing}
Checkpointing is a well-known technique that allows tasks to progress in spite of failures\cite{DBLP:journals/tse/KooT87}. When using checkpointing, snapshots of a task are taken periodically. In case of failure, a task can be restarted at its last snapshot instead of the start, saving a significant amount of time. However, checkpointing is not without costs. Creating a snapshot can take significant time depending on the type of task. Because of this, creating snapshots too frequently can result in the overhead of checkpoints exceeding its gains. In this section, we compare the impact of different checkpointing on the performance of different data centers exposed to different failure models.

\subsection{Experiment Setup}
Experiments are run using the OpenDC simulator\footnote{\url{opendc.org}}, an open source data center simulator. Using OpenDC, we are able to replay the same workload trace in combination with different failure traces and checkpointing models. 

\begin{table}[]
    \centering
    \caption{Workloads used in experiments.}
    \label{tab:checkpoint:workload_traces}
    \resizebox{\linewidth}{!}{
    \begin{tabular}{c|r|r|c}
    \hline
    \textbf{Workload} & \textbf{Duration} & \textbf{Number of tasks} &\textbf{ Average task duration} \\\hline
    Lisa   & 124 days & 194,917 & 01:49:38 \\
    Marconi    & 30 days & 22,938 &  06:20:12 \\
    \hline
    \end{tabular}}
\end{table}

\subsubsection{Workloads}
The executed workload can influence the impact of failures. To understand the impact of the workload trace, we use workload traces from two different compute clusters in our experiment, SURF Lisa~\cite{DBLP:journals/fgcs/VersluisCGLPCUI23} and CINECA Marconi~\cite{andrea_borghesi_2023_7590583}. The Lisa workload is 5 months long and is executed on a 150-node configuration. The Marconi workload is 1 month long and is executed on a 700-node configuration (see \autoref{tab:checkpoint:workload_traces}). 

\begin{table}[]
    \centering
    \caption{Failure traces used in experiments. TBF: average Time Between Failures, Duration: The average duration of a failure, Intensity: The average ratio of hosts that are effected by an failure.}
    \label{tab:checkpoint:failure_traces}
    \begin{tabular}{c|c|c|c}
    \hline
    \textbf{Failure trace} & \textbf{TBF} & \textbf{Duration} &\textbf{Intensity} \\\hline
    Facebook Massenger    & 2 days 15:50:22 & 00:36:51 & 0.299 \\
    Gmail    & 1 days 06:57:13 & 00:26:48 & 0.572 \\
    Whatsapp   & 0 days 14:35:48 & 00:26:44 & 0.579 \\
    \hline
    \end{tabular}
\end{table}

\subsubsection{Failure Traces}
Three different failure traces are used with a low (Whatsapp), medium (Gmail), and high (FaceBook Messager) frequency of failures (see \autoref{tab:checkpoint:failure_traces}). The starting point of the failure trace determines when a failure occurs and can thus influence the results. To combat this, each experiment is run five times, starting at different points in the failure trace. 

\subsubsection{Checkpointing models}
We compare the impact of six different checkpoint models that vary in the frequency of creating snapshots. The "no checkpoints" model does not make any checkpoints and is used as a baseline. The next four models use a static interval time between snapshots of 10 minutes, 1 hour, 10 hours, and 24 hours. Finally, the "adaptive" model uses a dynamic interval time between snapshots, starting at 10 minutes and increasing the interval by multiplying it by 1.5 after each snapshot. After a failure, the interval is restored to 10 minutes.

\begin{figure*}[t]
    \centering
     \begin{subfigure}[b]{\textwidth}
         \centering
         \includegraphics[width=\textwidth]{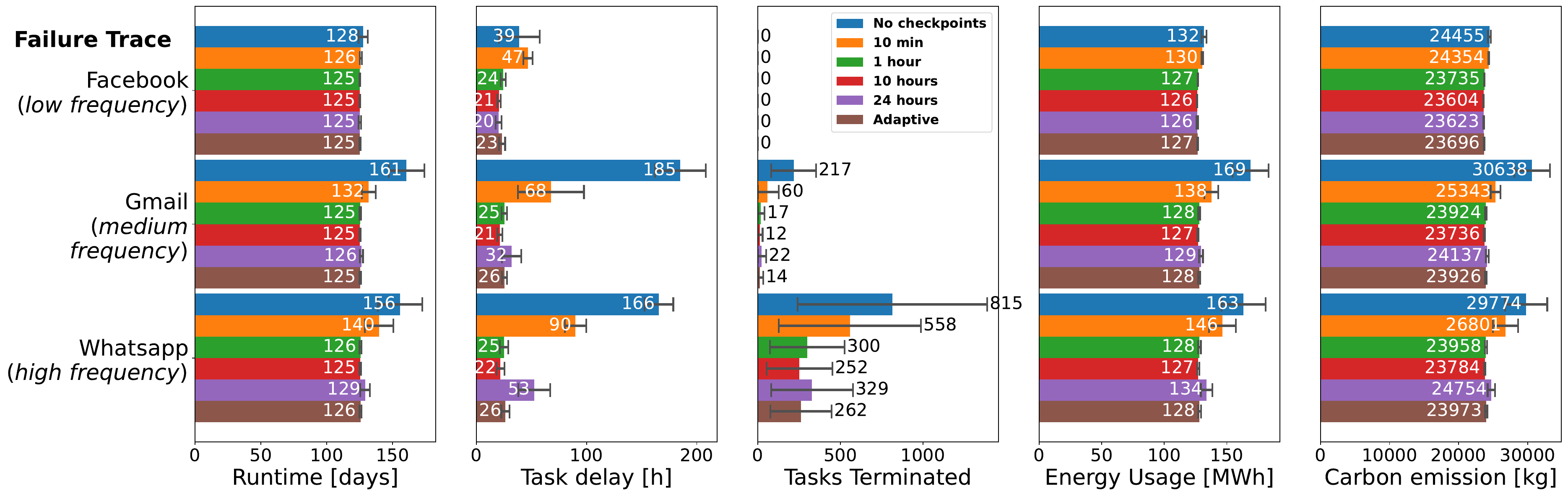}
         \caption{Workload Trace: SURF Lisa}
         \label{fig:checkpoint:resultsurfsara}
     \end{subfigure}
      \begin{subfigure}[b]{\textwidth}
         \centering
         \includegraphics[width=\textwidth]{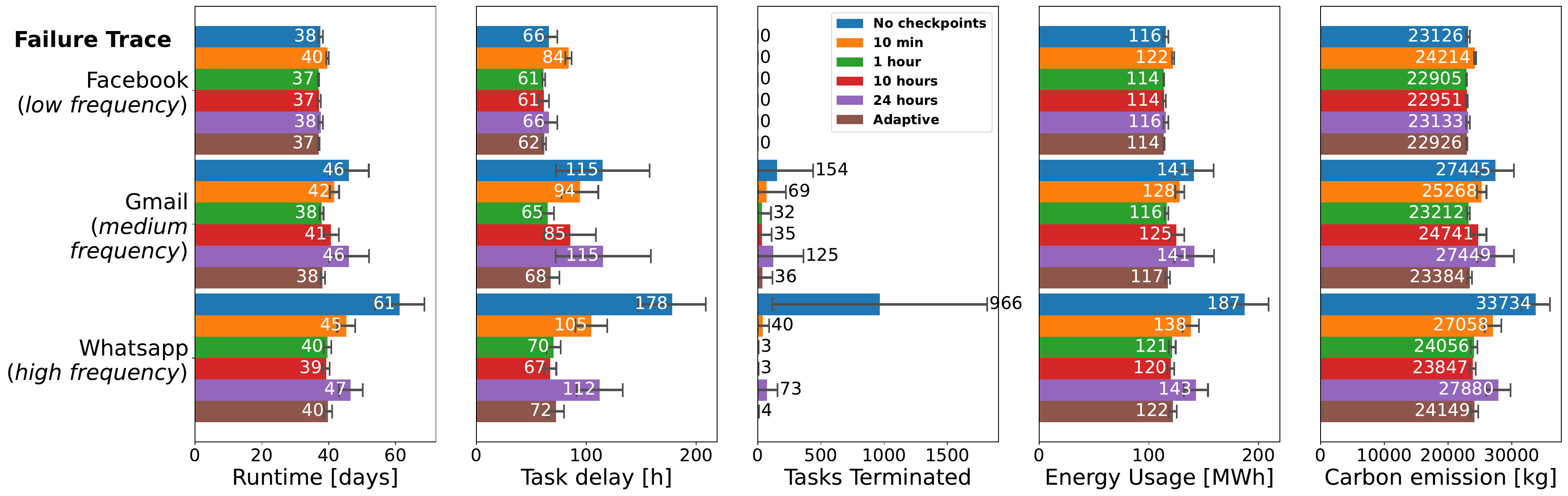}
         \caption{Workload Trace: CINECA Marconi}
         \label{fig:checkpoint:resultmarconi}
     \end{subfigure}
     \caption{Impact of checkpoint model and failure trace on different metrics for two worklaod traces.}
    \label{fig:checkpoint:resultsbymodel}
\end{figure*}

\subsection{Metrics}
The impact of checkpointing is determined using five metrics that focus on performance, quality of service, and sustainability.  

\subsubsection{Runtime}
The time required to complete all tasks in a workload. This is also often referred to as the makespan. 
\subsubsection{Task Delay}
The average delay of a task. This is calculated by subtracting the theoretical best execution time (the task is executed directly on submission and is never delayed) from the actual execution time of a task. This metric provides insight into the customer's quality of service.  
\subsubsection{Tasks Terminated}
The number of tasks that the data center terminated due to excessive failures. In our experiments, tasks are terminated by the data center when they fail more than 10 times. 
\subsubsection{Energy Usage}
The total energy used by the data center during the workload. Note: we assume that when a machine fails, it will keep using energy as if it is idle. 
\subsubsection{Carbon Emission}
The total carbon emitted during the workload caused by energy usage. Carbon emission is calculated using traces collected from ENTSO-E\footnote{\url{https://www.entsoe.eu/}}. More specifically, we are using carbon intensity traces from the Dutch power grid in the year 2022.

\subsection{Experiment Results}
\autoref{fig:checkpoint:resultsbymodel} depicts the impact of using different checkpointing models on the execution of the SURF Lisa (\ref{fig:checkpoint:resultsurfsara}) and the CINECA Marconi (\ref{fig:checkpoint:resultmarconi}) workload when exposed to different failure models. \\

\textit{The Impact Of Checkpointing:} When exposed to medium- and high-frequency failure traces, the use of checkpointing significantly affects the performance of both the SURF Lisa and the CINECA Marconi workloads (\refmainfinding{mf:checkpoint1}). The runtime of the Lisa workload is reduced by 22\% and 19\% when using the best checkpointing model for the medium- and high-frequency failure traces respectfully. Similarly, the runtime of the Marconi workload is reduced by 17\% and 35\%.  
Using checkpointing does not always improve performance. when exposed to a low-frequency failure trace, the runtime of the SURF Lisa workload is reduced by only 2\% when using the best checkpointing model. The only metric that is significantly affected by checkpointing is the average task delay, which is reduced by 48\% and 7\% for the Lisa and Marconi workload, respectively (\refmainfinding{mf:checkpoint3}). 
Using checkpointing can also result in worse performance. For example, when exposed to the low-frequency failure trace, the runtime of the Marconi workload increases by 5\% when using the "10 min" checkpointing model (\refmainfinding{mf:checkpoint2}).\\

\textit{Comparison Between Checkpointing Models:} Selecting the correct checkpointing model can significantly affect the performance of a data center. The best performing checkpointing models are "1 hour", "10 hours", and "adaptive", which alternate between being the best performing model depending on the specific experiment (\refmainfinding{mf:checkpoint4}). The "24 hours" model performs badly on the medium- and high-frequency failure traces, in which it does not have the time to create any checkpoints before failure. Finally, the "10 min" model is consistently significantly outperformed by at least three other models in every experiment. When exposed to the low-frequency failure trace, the "10 min" model can even result in worse performance compared to using no checkpointing. In these cases, the overhead of making frequent snapshots exceeds the gain on failure.\\

\textit{Comparison Between Different metrics:}
The impact of checkpointing is not the same for all metrics. Runtime, energy usage, and carbon emissions behave very similarly. This is no surprise given that a longer runtime is caused by an increased idle time, which results in higher energy usage. Higher energy usage, in turn, results in higher carbon emissions. In contrast, Task delay, and Task Terminated behave differently. In most experiments, the effect on Task Delay is similar to the other metrics but much more extreme. For example, the runtime of the SURF Lisa workload exposed to the high-frequency failure trace can be reduced by only 2\% when using the best checkpointing model. This same checkpointing model will reduce Task Delay by over 48\%. This shows that while failures and checkpointing might influence the overall performance of a data center, they can still significantly impact the customer(\refmainfinding{mf:checkpoint3}).

%% file: sections/8.experiment.tex
\section{Using Failure Traces to Estimate the Number of Retries in Service-based Applications}
\label{sec:retries}
\vspace{-0.5em}

We demonstrate how user-reported failure traces can be used in microservice-based computer systems experiments and how they impact the results. Our main finding is:

\begin{description}
    
    \defmainfinding{mf:10x-less}{A real trace leads to more than $10\times$ less retries, but $100\times$ more failures compared to a constant failure probability for applications composed of a long chain of services.}
    
\end{description}

Complex cloud applications such as e-commerce, social media and banking are important to daily life. Latency is an important metric for such applications. For example, increased latency reduces the click-through rate of web search~\cite{DBLP:conf/sigir/ArapakisBC14}. Complex cloud applications are composed of tens, or even hundreds of (micro-)services, working together to perform a task. The latency of the application depends on the latency of each of service. It also depends on how often the services fail. For example, additional latency appears when a service failure leads to a request being retried. Existing work~\cite{DBLP:conf/nsdi/PrimoracAB21, DBLP:conf/conext/VulimiriGMSRS13} considers \textit{constant failure probability} or similar simple models to estimate the additional latency. We show that using user-reported failure traces produces different results. Specifically, most requests succeed with fewer retries than if a constant failure probability was assumed, but during failure periods, a lot more requests fails.

We conduct experiments with multi-service apps, using trace-based simulation. We interpreting the number of user reports of a service as the probability that a request to the service fails. We scale the number of reports by the number of users of the service at that time. As we do not know the actual number of users, we assume that the number of users follows a diurnal pattern with peak usage in the evening; we approximate this with a shifted sine wave. After scaling the trace, we normalize it into probabilities [0, 1] by dividing each data point by the maximum number of user reports in a single time period (20 minutes in our case). We thus obtain a failure probability associated with every 20-minute period in the trace.

We choose three app structures---for each, a number of microservices and their interdependencies. Many microservice applications are composed of simple structures~\cite{DBLP:journals/software/EismannSESGHAI21} such as (1)~monolith (a single microservice called by a client), (2)~fanout (a client makes multiple requests to a microservice all of which need to succeed), and (3)~long chain (a client calls a microservice, which calls another, and so in a long chain). We simulate failures in each of these structures and measure the impact on latency.

We use the metric number of retries as an indicator of the tail latency experienced by the service. Tail-latency is particularly important in cloud operations~\cite{tail_at_scale}, where services are possibly used by millions of users daily. We assume each microservice and client comes with a retry oracle which knows when to retry. Hedging~\cite{DBLP:conf/nsdi/PrimoracAB21} and failure detection algorithms~\cite{DBLP:journals/csur/FreilingGK11} have been studied extensively, and our results complement that work. We also assume that cost of retry dominates the latency, compared to the actual request-processing time; this holds for short-running apps~\cite{DBLP:conf/usenix/ShahradFGCBCLTR20}.

\begin{figure}[t]
    \centering
    \includegraphics[width=0.48\textwidth]{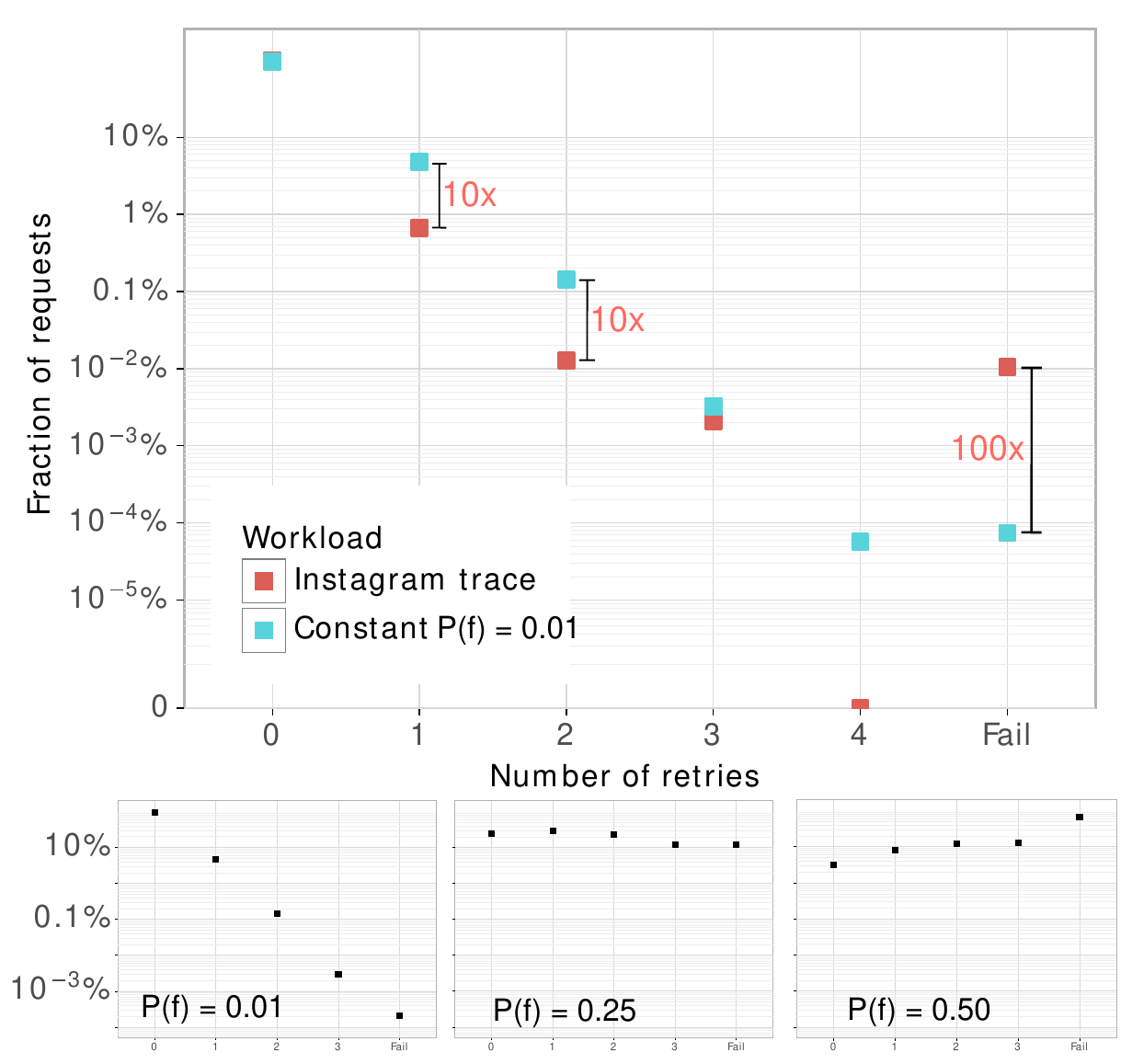}
    \caption{Distribution of retries for a long chain chain microservice structure with 5 microservices. At the bottom are the distributions at different failure probabilities indicating why the distribution for the trace looks like it does. In particular, notice the large number of failed requests at high failure probability and for the trace. The vertical axis is logarithmic.}
    \label{fig:long_chain}
\end{figure}

We evaluate the number of retries experienced by a long chain workload structure comprised of 5 microservices. The evaluation for other structures is in the technical report. At each stage, we allow a maximum of 3 retries, which is common in practise. Reaching the maximum number of retries at a stage means that the previous stage has to retry again. When the retries of the first stage are exhausted, the request is considered to have \textit{failed}. Figure~\ref{fig:long_chain} depicts the results. The horizontal axis represent the number of retries with a special marker for failed requests. The vertical axis is logarithmic axis depicting the fraction of requests. A point on the plot represents the fraction of requests that required a certain amount of retries to succeed. We run the experiment with two different workloads: (new) the Instagram use-reported failure trace and (traditional) a constant failure trace. The constant failure probability we chose is 1\%~(P(f) = 0.01). The arrival pattern is diurnal, peaking in the evening; we approximate this with a shifted sine wave.

We find that the number of requests that succeed is, in total across all possible retry-counts and also individually for retry-counts 1 and 2, more than  $10\times$ lower for the Instagram trace compared to the constant failure probability~(\refmainfinding{mf:10x-less}). Requests which require 4 retries are non-existent for the Instagram trace, but they appear prominently with constant failure probability. Significant differences appear also in failed requests; $100\times$ more requests fail with the Instagram trace compared to the constant failure probability. The number of requests which succeed with 3 retries is very close for both the Instagram trace and constant failure probability. To understand the unique shape of the failure distribution exhibited when using the Instagram trace, we plot the retry distribution with different failure probabilities found in the trace at the bottom of Figure~\ref{fig:long_chain}. Notice the flat line at 25\% failure probability, which is the likely cause of the high number of requests that succeed with 3 retries for the Instagram trace. Similarly, at 50\% failure probability, the number of failed requests becomes very high. Such periods with high failure probability lead the high number of failed requests for the Instagram trace.

%% file: sections/9.conclusion.tex
\section{Discussion and Threats to Validity}\label{sec:discussion}

\textit{Data Accuracy:} We collect data from different sources such as operator status pages and user-report crowdsourcing websites. Each of these sources has its interpretation of failure, making it challenging to evaluate the accuracy of data with respect to the ground truth. We compare different failure report sources for the same service in \Cref{sec:validation} as a sanity check to ensure that our data is accurate at least when it comes to major failures.

\textit{Crowdsourced Data Collection:} Lack of data from multiple crowdsourced data sources biases our work to the pool of users who report to the particular crowdsourcing services we collected the data from, Outage Report and DownRightNow. It also exposes the number of user reports to vagaries such as changing search-engine ranking (i.e., when Outage Report is not the first result), changing social media popularity of the service, and other causes we cannot control. For example, the reduction in number of user reports could be due to the reduced popularity of the service and not to fewer actual failures.

\textit{Comparison to Old Datasets:} We compare our newly collected data to a previous dataset from the Failure Trace Archive (FTA)~\cite{DBLP:journals/jpdc/JavadiKIE13}. The FTA dataset was collected by visiting different websites periodically from a central server. That is a different method from using crowdsourced and operator reports. This difference makes high-precision numerical comparisons between the datasets error-prone. However, we believe our findings hold because the failure during and time between failures of the two datasets are at least an order of magnitude different.

\textit{Comparison Between Different Services:} We know that comparing multiple services and cloud operators is a non-trivial pursuit. Our selective experiments and findings should not be interpreted as stating conclusively that one service or operator is better than the others. We do not have sufficient data to investigate such claims. Such data could arise from a community-wide study on multi-year data and would be facilitated by the release of ground-truth data by the cloud operators. Our study aims to take the first step toward a systematic data collection framework, identify patterns, and quantify the bounds of typical failures. Such findings across broad categories such as web, cloud, and online games from both user and operator perspectives are of interest to clients and infrastructure planners for building reliable services.

\begin{table*}[t]
\setlength{\tabcolsep}{3pt}
\centering
 \small
    \caption{Overview of related work. (\textbf{TW} = This Work)} 
    \label{tab:related_work}
    \centering
    \begin{tabular}{clp{7cm}l}
        \toprule
        \textbf{Feature} & \textbf{Type} & \textbf{Articles} & \textbf{Comments} \\
        \midrule
        \multirow{5}{*}{\rotatebox[origin=c]{90}{Data Source}} & Crowdsourced & \textbf{TW-only} & Users' experience with cloud services.\\
        & Cloud Providers & \cite{DBLP:conf/cloud/GunawiHSLSAE16} \& \textbf{TW} & Data available from cloud provider or the news.\\
        & Multiple Clusters & \cite{DBLP:conf/icdcs/El-SayedZS17, DBLP:journals/jpdc/JavadiKIE13, DBLP:conf/sc/GuptaPET17, DBLP:conf/dsn/BirkeGCWE14, DBLP:conf/dsn/TaherinPGLT21, DBLP:conf/dsn/KumarJMKHSKKIB20}& Data for multi-cluster infrastructure.\\
        & Single Cluster & \cite{DBLP:conf/sosp/ChenSGK11, DBLP:conf/ccgrid/KavulyaTGN10, DBLP:conf/ipps/ZhengYTLGDCB11, DBLP:conf/dsn/El-SayedS13, DBLP:conf/dsn/MartinoKIBFK14, DBLP:conf/dsn/RosaCB15, DBLP:conf/dsn/GuptaTJRM15, DBLP:conf/dsn/DiGPSC19}& Data for infrastructure of a single cluster.\\
        & User Reports & \cite{DBLP:conf/osdi/YuanLZRZZJS14, DBLP:conf/sosp/YinMZZBP11, DBLP:conf/cloud/GunawiHLPDAELLM14, DBLP:conf/dsn/FonsecaLSR10, DBLP:conf/dsn/FrattiniGCRT13, DBLP:conf/cloud/GunawiHSLSAE16}& Bug reports for individual software. \\
        \midrule
        \multirow{4}{*}{\rotatebox[origin=c]{90}{Workload}} & Cloud & \cite{DBLP:conf/dsn/FrattiniGCRT13, DBLP:conf/dsn/RosaCB15, DBLP:conf/cloud/GunawiHSLSAE16} \& \textbf{TW}& Data from cloud datacenters or cloud applications. \\
        & HPC & \cite{DBLP:conf/icdcs/El-SayedZS17, DBLP:journals/jpdc/JavadiKIE13, DBLP:conf/ipps/ZhengYTLGDCB11, DBLP:conf/sc/GuptaPET17, DBLP:conf/dsn/El-SayedS13, DBLP:conf/dsn/MartinoKIBFK14, DBLP:conf/dsn/GuptaTJRM15, DBLP:conf/dsn/DiGPSC19, DBLP:conf/dsn/TaherinPGLT21, DBLP:conf/dsn/KumarJMKHSKKIB20}& Supercomputer applications and hardware. \\
        & Big Data & \cite{DBLP:conf/osdi/YuanLZRZZJS14, DBLP:conf/icdcs/El-SayedZS17, DBLP:conf/ccgrid/KavulyaTGN10, DBLP:conf/cloud/GunawiHLPDAELLM14}& Data processing systems like MapReduce. \\
        & Enterprise & \cite{DBLP:conf/sosp/YinMZZBP11, DBLP:conf/sosp/ChenSGK11, DBLP:journals/jpdc/JavadiKIE13, DBLP:conf/dsn/FonsecaLSR10, DBLP:conf/dsn/BirkeGCWE14}& From enterprise vendors such as IBM. \\
        \midrule
        \multirow{3}{*}{\rotatebox[origin=c]{90}{Analysis}} & \multirow{1}{*}{Distribution} & \cite{DBLP:conf/icdcs/El-SayedZS17, DBLP:conf/sosp/ChenSGK11, DBLP:conf/ccgrid/KavulyaTGN10, DBLP:journals/jpdc/JavadiKIE13, DBLP:conf/ipps/ZhengYTLGDCB11, DBLP:conf/cloud/GunawiHLPDAELLM14, DBLP:conf/sc/GuptaPET17, DBLP:conf/dsn/El-SayedS13, DBLP:conf/dsn/BirkeGCWE14, DBLP:conf/dsn/RosaCB15, DBLP:conf/dsn/GuptaTJRM15, DBLP:conf/dsn/DiGPSC19, DBLP:conf/cloud/GunawiHSLSAE16, DBLP:conf/dsn/TaherinPGLT21, DBLP:conf/dsn/KumarJMKHSKKIB20} \& \textbf{TW}& \multirow{1}{*}{Report empirical distributions of failure features.} \\

        & \multirow{1}{*}{Trend} & \cite{DBLP:conf/sosp/ChenSGK11, DBLP:conf/ccgrid/KavulyaTGN10, DBLP:conf/ipps/ZhengYTLGDCB11, DBLP:conf/sc/GuptaPET17, DBLP:conf/dsn/El-SayedS13, DBLP:conf/dsn/FonsecaLSR10, DBLP:conf/dsn/FrattiniGCRT13, DBLP:conf/dsn/MartinoKIBFK14, DBLP:conf/dsn/GuptaTJRM15, DBLP:conf/cloud/GunawiHSLSAE16, DBLP:conf/dsn/TaherinPGLT21, DBLP:conf/dsn/KumarJMKHSKKIB20} \& \textbf{TW}& \multirow{1}{*}{Report trends over time, location, lifetime, etc.} \\
        
        & \multirow{1}{*}{Categorization} & \cite{DBLP:conf/icdcs/El-SayedZS17, DBLP:conf/osdi/YuanLZRZZJS14, DBLP:conf/sosp/YinMZZBP11, DBLP:conf/sosp/ChenSGK11, DBLP:conf/ccgrid/KavulyaTGN10, DBLP:conf/cloud/GunawiHLPDAELLM14, DBLP:conf/sc/GuptaPET17, DBLP:conf/dsn/FonsecaLSR10, DBLP:conf/dsn/FrattiniGCRT13, DBLP:conf/dsn/MartinoKIBFK14, DBLP:conf/dsn/BirkeGCWE14, DBLP:conf/dsn/RosaCB15, DBLP:conf/dsn/DiGPSC19, DBLP:conf/cloud/GunawiHSLSAE16, DBLP:conf/dsn/TaherinPGLT21, DBLP:conf/dsn/KumarJMKHSKKIB20} \& \textbf{TW}& Categorize failures based on severity, root cause, etc.\\
        \bottomrule
    \end{tabular}
\end{table*}

\section{Related Work}
\label{sec:related_work}

Understanding failures is a long-lasting area of interest in computer systems research.
We conduct a systematic literature survey~\cite{kitchenham2007guidelines} of failure studies related to distributed systems
for the articles published in top conferences and journals between 2010 and 2024.
Table~\ref{tab:related_work} summarizes the 21 articles we found closest to our failure characterization; although only 3 focus on clouds, the others characterize failures in large-scale systems.
The mapping of our study to the table emphasizes the novel aspects of our work: unique data and an unique mix of approaches for analysis, for a cloud workload; this leads to new findings.

Closest to our is the Failure Trace Archive~\cite{DBLP:journals/jpdc/JavadiKIE13} which collected data about the availability of networks of workstations, websites, and some HPC clusters till~2009. Our data about web and cloud service is qualitatively and quantitatively different. We collect and analyze data at the service level, a higher level abstraction than hardware. The world has also changed significantly, with online services involved in every part of life. Networks of workstations and HPC clusters are not involved in daily life to the same extent. The quantitative differences are analyzed in \Cref{sec:tbf} and \Cref{sec:patterns}.

Second closest to our work, Gunawi et al.~\cite{DBLP:conf/cloud/GunawiHSLSAE16}  analyze cloud failures over a multi-year period, using as input data news-reports.
Complementary to our time-related analysis, they find that the distribution of failure duration in a bad year skews higher than in an average year. 
Most of their work focuses on the root-cause analysis, which is out of scope for this work.

\textit{Failure Characterization Studies:} User-reported bugs for systems software (e.g., Hadoop, Cassandra) have been widely studied. 
Although tools for collecting and analyzing the client-side failure behavior of cloud services exist~\cite{DBLP:conf/nsdi/BurnettCCEGJMPR20}, how users experience the behavior of cloud services is still virtually unknown. 
This work aims to address this knowledge gap.

Other characterizations of failures in cloud computing analyze environments different from our work:
a single virtual environment~\cite{DBLP:conf/dsn/FrattiniGCRT13},
a cluster a Google running big data applications~\cite{DBLP:conf/dsn/RosaCB15}.
These studies are very similar in nature to the specific studies focusing on grid computing and peer-to-peer~\cite{DBLP:journals/jpdc/JavadiKIE13}, and similar environments of the 2000s.
All are different from our client-centric analysis of end-user experience of cloud providers and SaaS services.

We compare our work to other studies on multi-cluster environments. 
Several studies observe the presence of short and/or long interarrival times~\cite{DBLP:conf/dsn/BirkeGCWE14},
bursty and correlated failures~\cite{DBLP:journals/jpdc/JavadiKIE13},
time-patterns over days of the week and months of the year~\cite{DBLP:conf/sc/GuptaPET17},
etc.
But our study complements this body of work: the quantitative results we present are often different, e.g., minute-length failures for most SaaS services, in contrast to the hours to days observed for multi-clusters.

\textit{Trace Archives and Datasets:} Multiple trace archives have been created to benefit the community. Example of trace archives include parallel workloads archive~\cite{DBLP:journals/jpdc/FeitelsonTK14}, grid workloads archive~\cite{DBLP:journals/fgcs/IosupLJADWE08}, cloud software bugs~\cite{DBLP:conf/cloud/GunawiHLPDAELLM14}, workflow trace archive~\cite{DBLP:journals/tpds/VersluisMTHPDI20}, and SNIA IOTTA~\cite{snia_iotta}. Multiple datasets have also been published independently to benefit the community such as the Supercloud dataset~\cite{DBLP:conf/hpec/SamsiWBLJREABHH21}, Twitter KV-traces~\cite{DBLP:journals/tos/YangYR21}, CINECA Marconi dataset~\cite{borghesi2023m100}, and HPC-ML workloads~\cite{chu2024generic}. Cloud providers like Google~\cite{clusterdata:Wilkes2020}, Microsoft~\cite{DBLP:conf/osdi/HadaryMMPGDDJCR20}, and Alibaba~\cite{DBLP:conf/nsdi/WengXYWWHLZLD22} have also released cluster usage traces. All these traces have been used in thousands of papers and have pushed the field forward.


\section{Conclusion}\label{sec:conclusion}

In digital societies, cloud services are pervasive in every industry and affect most people. However, we do not understand their reliability well, not as well as we do for, e.g., cars. We propose collecting data about the availability of popular online services as the first step towards solving this problem.

We collect long-term data about failures of online services such as infrastructure clouds (AWS, Azure, GCP), online services (Github, Atlassian, Facebook, Netflix, etc.), and online games (Runescape, Minecraft). We collect this data from two vantage points, the service operators and the users. We summarize data and their sources in \Cref{tab:dataset_summary}.

We demonstrate the structure and basic utility of this data in \Cref{sec:anatomyofafailure}. We describe our data collection, cleaning, interpretation, and analysis process in \Cref{sec:method}. We also describe our trace format. We validate the data by comparing the data for the same failures from operator reports and user reports for two services, Github and AWS, in \Cref{sec:validation}.

We analyze this data through a standard MTBF and MTTR reliability analysis in \Cref{sec:tbf}, time patterns in \Cref{sec:patterns}, failure severity in \Cref{sec:severity}, user-reported symptoms, and operator reported symptoms in \Cref{sec:causes}. This analysis results in 14~observations.

We demonstrate the utility of the failure traces we collect by evaluating the impact of traces on checkpointing policies in \Cref{sec:checkpoint} and retry policies in \Cref{sec:retries}. This simulator-based experimental analysis leads to 5~observations.

Finally, we release the failure data, analysis code, and simulation code as open-source artifacts at \url{https://github.com/atlarge-research/cloud-uptime-archive}. The dataset is also available on Zenodo at \url{https://doi.org/10.5281/zenodo.14712441}. We call the community to enhance this dataset with new failure data sources, analysis techniques, and using the traces in the evaluation of future fault-tolerant systems.

\section*{Acknowledgements}

This work is supported by EU Horizon Graph Massivizer and EU MSCA CloudStars projects. This research is partly supported by a National Growth Fund through the Dutch 6G flagship project “Future Network Services”.